\documentclass[twocolumn]{aastex61}

\shorttitle{Deadly cusps: Eridanus II and Andromeda XXV}
\shortauthors{Nicola C. Amorisco}

\begin{document}

\title{Deadly dark matter cusps vs faint and extended star clusters: \\Eridanus~II and Andromeda~XXV}

\email{nicola.amorisco@cfa.harvard.edu}

\author{Nicola C. Amorisco}
\affil{Institute for Theory and Computation,  Harvard-Smithsonian Center for Astrophysics,  60 Garden St., Cambridge,  MA 02138,  USA}
\affil{Max Planck Institute for Astrophysics,  Karl-Schwarzschild-Strasse 1,  D-85740 Garching,  Germany}



\begin{abstract}
The recent detection of two faint and extended star clusters in the central regions of two Local Group 
dwarf galaxies, Eridanus~II and Andromeda~XXV, raises the question of whether clusters 
with such low densities can survive the tidal field of cold dark matter haloes with central density cusps. 
Using both analytic arguments and a suite of collisionless N-body simulations, 
I show that these clusters are extremely fragile
and quickly disrupted in the presence of central cusps $\rho\sim r^{-\alpha}$ with $\alpha\gtrsim 0.2$.
Furthermore, the scenario in which the clusters were originally more massive and sank to the center
of the halo requires extreme fine tuning and does not naturally reproduce the observed systems. 
In turn, these clusters are long lived in cored haloes, whose central regions are safe shelters for $\alpha\lesssim 0.2$. 
The only viable scenario for hosts that have preserved their primoridal cusp to the present time 
is that the clusters formed at rest at the bottom of the potential, which is easily tested by 
measurement of the clusters proper velocity within the host. This offers means 
to readily probe the central density profile of two dwarf galaxies as faint as 
$L_V\sim5\times 10^5 L_\odot$ and $L_V\sim6\times10^4 L_\odot$, in which stellar feedback is
unlikely to be effective.

\end{abstract}




\section{Introduction} 

The distribution of matter on cosmological scales is
very successfully reproduced by the standard $\Lambda$ cold dark matter (DM) model:
the agreement with measurements of both the cosmic microwave background and the baryonic acoustic oscillation feature 
\citep[e.g.,][]{Planck,BAO,FW12} is impressive. 
However, these tests only probe the DM linear power spectrum at scales $\gtrsim$10 Mpc. 
At the scales of galaxies and below alternative DM
models make different predictions, which provides means to differentiate among them.

Cold, non relativistic DM particles virialize in haloes characterized 
by a central density distribution which diverges as $\rho\sim r^{-1}$ \citep{DC91,NFW96}, 
and containing a fraction of about 10\% of their mass in substructure, in the form of 
bound sub-haloes \citep{JD08,VS08}.
Warm(-er) DM particles allow for less power at small scales: the subhalo mass function 
is suppressed below some model-dependent minimum mass and the total mass fraction in substructure
is lowered \citep[e.g.,][]{Bo01,Men12,ML14,SB16}; halo concentration, additionally, becomes a 
non-monotonic function of halo mass \citep[e.g.,][]{AL16}.
Models allowing for self-interactions also imply a reduction in the small-scale power,
but additionally feature haloes with centrally cored density profiles \citep[e.g.,][]{SS00,VZ12,El15}, 
with the size of the core depending on the strength of the interaction itself \citep{JZ13,LA16}.
Central density cores are also predicted in the `fuzzy' DM scenario \citep{Pr90,Hu00,Hui17},
in which DM is made of light scalar particles that manifest their quantum properties at
astrophysical scales. Cores sizes are dictated by the mass of the axion-particle
and recent numerical studies are beginning to provide definite predictions for the process 
of cosmological structure formation within this model \citep{HS14,HS16,PM15,Du17}.

Establishing sound astrophysical tests for dark matter models on dwarf galaxy scales 
and below has proven especially hard so far. 
Recently, probing the mass function of halo substructure, with either strong lensing \citep[e.g.][]{KD04,SV14,YH16}
or thin stellar streams \citep[e.g.,][]{Yo11,RI02,DE16} is establishing itself as a promising venue. 
Here, however, I concentrate on those complimentary tests based on the detailed properties of the density 
profile of low-mass, dark matter dominated galaxies (virial mass $M_{200}\lesssim10^{11} M_{\odot}$).

On the theoretical side, the predictions for the halo density profile mentioned above do not take into 
account the impact of baryons, and the complex hydrodynamical processes that accompany galaxy formation. 
Radiation and winds from young stars and supernovae, often globally referred to as stellar feedback,
are an important ingredient in the formation of dwarf galaxies \citep[e.g.,][]{DS86,JN96,Ma06,PG14}.
However, a consensus has yet to be reached on how feedback can or cannot sculpt cores into the central regions of 
originally cuspy dwarf galaxy haloes \citep[e.g.,][]{FG12,AZ12,KE16,TS16,AF16}.
{Simple} energetic arguments \citep{JP12, NA14} and some sub-grid implementations
of the feedback processes in hydrodynamical simulations \citep[e.g.,][]{DC14,JO15} suggest that core-creation 
is suppressed in faint enough galaxies ($L\lesssim10^{5.5} L_\odot$). Dwarfs with these luminosities should
preserve their primordial cusps and represent perfect targets to test the nature of dark matter.
However, a different numerical implementation \citep[][]{JR16} suggests that  
cores can emerge also in the faintest galaxies. 
This motivates even more strongly the need for reliable measurements of the inner density profile of low mass haloes. 
Even before probing the nature of DM, these measurements are crucial to understand the 
feedback processes themselves.
 
On the observational side, however, such measurements remain extremely challenging. 
Decades of debate have shown that this is the case for galaxies supported 
by rotation \citep[e.g.,][]{PS91,FP94,dB08,Oh11,JA14,KO15}. In addition, systems 
that are faint enough to have possibly preserved a pristine cusp are pressure supported systems.
Close enough dwarfs are the satellite galaxies of the Local Group; these include the `classical' dwarf Spheroidals 
(dSphs, $L\gtrsim10^5 L_\odot$) and the Ultrafaints \citep[UFs, $L\lesssim10^5 L_\odot$, e.g.,][]{MC12}. 
As for all pressure supported systems, their kinematic modelling is plagued by marked degeneracies,
which manifest themselves when line-of-sight kinematics alone is available. These degeneracies
make it impossible to measure the galaxy's density profile, and only allow for the determination of a mass 
scale \citep{MW09,JW10,NA11,AA14}, substantially complicating the analysis of datasets collected 
with painstaking effort \citep[e.g.,][]{GB08,MW09a}. This difficulty was partially overcome by the realization that 
chemo-dynamically distinct stellar subpopulations {can occur} in Local Group dSphs \citep[e.g.,][]{ET04,GB06,WP11,GK16} {and}
can be used jointly to constrain the gravitational potential in which all stars reside.
This kind of analysis can break the degeneracy between mass and anisotropy typical of pressure supported systems.
Two classical dSphs, Sculptor and Fornax, could be studied with this technique. In both cases, this 
is found to {weakly} disfavor a $\rho\sim r^{-1}$ cusp \citep{GB08,WP11,NA12,AA12,NA13,Zhu16}, 
but the statistical significance of this result has been contended \citep[e.g.][]{MB13,RF14,St17}.
Alternative methods proposed in the literature to probe the central dark matter profile of local dwarf galaxies include: 
(i) the survival time of unbound kinematic substructure \citep{Kl03,SS10};
(ii) the dynamical friction timescale of massive Globular Clusters \citep[e.g.,][]{HG98,SS06,Go06,Co12};
(iii) the internal kinematics of dwarf galaxy streams \citep{RE15}; 
(iv) the survival of loosely bound binary stars \citep{JP16}.
 
Motivated by the recent discovery of extended, low-mass star clusters in two Local Group dwarfs,
in this paper I seek to establish what are the constraints that their survival to the present day puts 
on the DM profile of their host haloes. Extended clusters have been observed before in
M31 \citep[e.g.,][]{Hux11}, where they are though to have been deposited by disrupted dwarfs
\citep[e.g.,][]{AM10,HM10,Hux13}. {The two recently discovered systems
are particularly extreme: as I will show, the combination of their stellar mass and size, together with the   
short orbital times and current projected locations make for an almost inescapable threat to their survival.
This represents the main difference from a previous study, \citet{JP09},  
dedicated to the dynamical evolution of the star clusters of the Fornax and Sagittarius dSphs.
Together with dense clusters, both these dwarf galaxies currently harbor diffuse star clusters, namely F1 in Fornax, Arp2 and Ter8 in Sagittarius. 
As shown by \citet{JP09}, these would easily be disrupted by the tides 
if they were to orbit close enough to the center. However, such fragile GCs 
are observed to lie at significant projected distances, where at the same time
(i) they are currently safe and (ii) dynamical friction is inefficient.
As a consequence, as recognized by \citet{JP09}, they are long lived.
This is not the case for the systems I explore in this paper.}

The {dwarf galaxies considered here} are Eridanus II 
\citep[EriII, ][]{SK15,Be15} in the periphery of the Milky Way, and Andromeda XXV 
\citep[AndXXV, ][]{Ri11, FC16}, around M31. Each of them contains an extended, low mass star cluster which,
given its structural properties, is extremely susceptible to the tidal field.
Still, in both systems, the cluster is observed to reside -- in projection -- in the central regions of the galaxy, 
where the tides are strongest in cold DM haloes.
Using both analytical arguments and collisionless numerical simulations I systematically explore 
the evolution scenarios that could allow the extended star clusters in EriII and AndXXV to survive 
to the present day.
Predictions for the internal kinematics of the clusters are also discussed, providing means to 
to distinguish between the different scenarios, and therefore to infer the density profiles of the two host galaxies.
Section 2 presents the two systems;
Section 3 {puts the dynamical problem into context}; 
Section 4 describes the numerical setup;
Section 5 describes possible scenarios for cuspy haloes;
Section 6 concentrates on cored haloes;
Section 7 discusses results and presents the Conclusions.

\section{The clusters and their hosts}

\subsection{Eridanus II}

At a distance of $D\approx370$ kpc, EriII is a UF satellite of the MW, with a luminosity of $L_V\approx 6\times10^4 L_\odot$,
a projected half-light radius of $R_h \approx$280~pc, a quite high ellipticity $\epsilon\approx 0.48$ and no evidence for the 
presence of gas \citep{DC16}. Using Magellan/IMACS spectroscopy \citet{Li17} have recently targeted EriII and confirmed 28 member stars. 
They find a velocity dispersion of $\sigma=6.9^{+1.2}_{-0.9}$~kms$^{-1}$, a mean metallicity of $\left[{\rm Fe/H}\right] \approx -2.4$,
and can exclude the presence of young stars in the system.

A round overdensity of stars near the center of EriII was already spotted in the discovery data \citep{SK15}, and subsequent deep
imaging \citep{DC16} has confirmed the presence of an extremely faint and quite extended star cluster. This has a luminosity of 
 \begin{equation}
L_{V,c}\approx 2\times10^3 L_\odot
 \label{lumE}
 \end{equation}
and a projected half-light radius of 
 \begin{equation}
R_{h,c}\approx13~{\rm pc}\ .
 \label{RhE}
 \end{equation}
As such, EriII is the least luminous
galaxy known to posses a stellar cluster, which has prompted the suggestions that other distant Milky Way GCs might in fact be hosted by
yet undiscovered low surface brightness galaxies \citep{DZ16}. 
The stellar clusters is projected very close to the inferred center of the stellar distribution
of EriII, but has a measurable offset of $d\approx 45$~pc. The resolved stars in the cluster would suggest 
its stellar population is old, and consistent with the old population of EriII ($\approx10$ Gyr).

\subsection{Andromeda XXV}

AndXXV sits at a projected distance of $\approx 90$~kpc from the center of M31 and has a luminosity of 
$L_V\approx 5\times10^5 L_{\odot}$ \citep{Ri11,MC12,NM16}, {approaching} the lower edge of the {range} 
of `classical' dSphs. Its projected half light radius is 
quite large for its luminosity, but also quite uncertain as a consequence of a chip gap in the available imaging data 
\citep{Ri11,NM16}, with literature values ranging 
between 550 and 700~pc. The internal kinematics of AndXXV was probed by \citet{MC13}, 
who measured a peculiarly low velocity dispersion of $\sigma=3.1^{+1.2}_{-1.1}$~kms$^{-1}$.

By visual inspection of stacked images, \citet{FC16} have recently unveiled the presence of a 
concentration of stars
near the central regions of AndXXV. The {color magnitude diagram} of the few 
resolved stars is compatible with that of old stars at the distance of AndXXV, which {suggests} 
that the cluster and the dwarf are indeed physically associated \citep{FC16}.
{Further support for this can be obtained by considering the probability of chance alignments, of either 
a foreground GC belonging to the MW or of a GC associated with M31. The first is very small: even assuming  
the MW has has many as 500 (yet undetected) GCs, distributed isotropically, the probability of a chance projection within an angle of $\approx 2 R_h/750$~kpc 
is of only\footnote{A similarly small probability is obtained using the same argument for the case of Eri~II.} 
$\approx3\times 10^{-4}$. The total number of GCs in M31 is uncertain, probably as high as $\sim450$ \citep[][]{Ve13},
 with as many as $\sim1000$ candidates \citep{Ga04}. If I assume the latter figure and a 3-dimensional 
number density profile that declines as slowly as $r^{-2}$ (truncated for example at $r=200$~kpc), the probability of 
 a chance alignment is of about $5\%$. This indicates that the cluster and AndXXV are very likely associated with each other.} 
 
The cluster has a luminosity of 
 \begin{equation}
L_{V,c}\approx 8\times10^3 L_\odot
 \label{lumA}
 \end{equation}
and a projected half light radius of 
 \begin{equation}
R_{h,c}\approx25~{\rm pc}\ ,
 \label{RhE}
 \end{equation}
making it significantly extended.
As for EriII, the projected position of the cluster does not coincide with the centroid of AndXXV main stellar component.
The precise value of the offset is uncertain due to the mentioned chip gap, but appears comparable to what {is} seen in
EriII, $d\approx 46$~pc.

\begin{figure*}
\centering
\includegraphics[width=\textwidth]{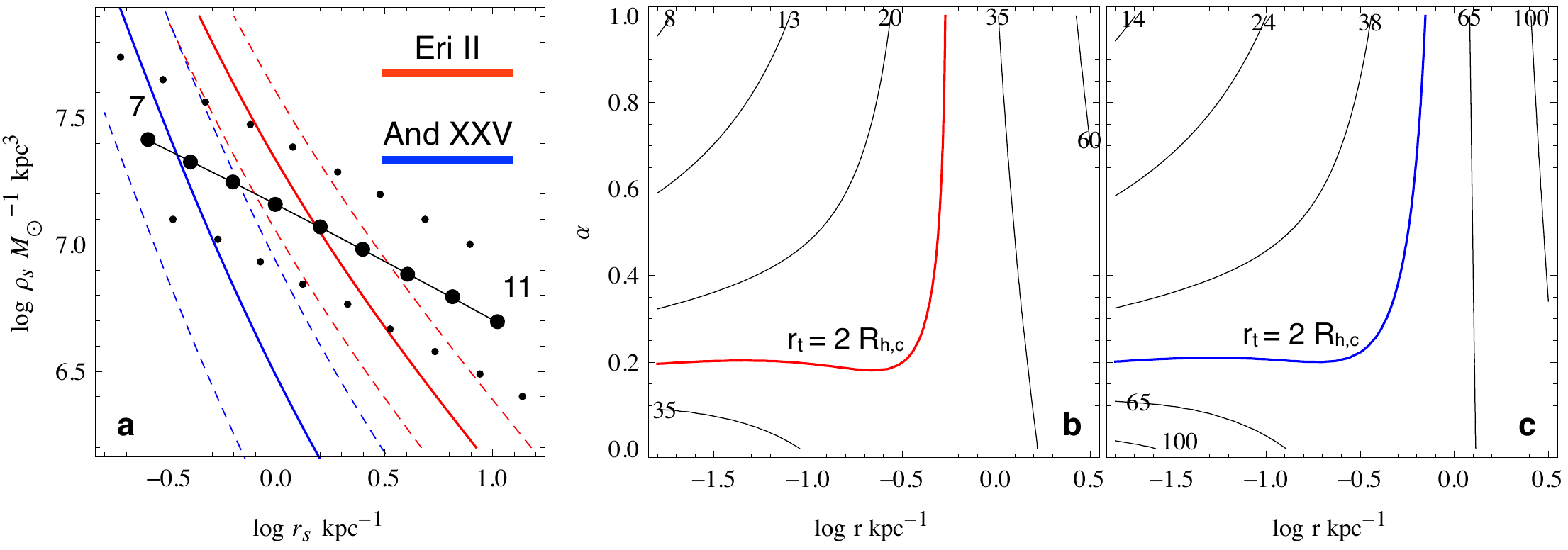}
\caption{Tidal radii for the star clusters observed in EriII and AndXXV. 
Panel $a$ shows the plane of the characteristic density $\rho_s$ and scale radius $r_s$
of an NFW density profile, and the 1-sigma constraints imposed by the observed stellar kinematics
in EriII (red lines) and in AndXXV (blue lines). Large points identify the mean properties of cosmological
cold dark matter haloes at redshift z=0, for virial masses between $10^7 M_\odot$ and $10^{11} M_\odot$,
in steps of 0.5 dex. Smaller black dots illustrate the size of the 1-sigma scatter in the mass-concentration relation.
Panels $b$ and $c$ show contours for the instantaneous tidal radius~(\ref{tidr}) in the plane of the 
galactocentric orbital distance $r$ and of the slope of the density profile $\alpha$ (see text 
for more details). Panel $b$ refers to a cluster with mass $M_c=5\times 10^3 M_\odot$, 
representative of the case of EriII, while panel $c$ is tailored on AndXXV, with $M_c=25\times 10^3 M_\odot$.     }
\end{figure*}

\section{Dynamical ingredients}


\subsection{Limits from the {dwarf galaxy} kinematics}

The observed values of the stellar velocity dispersion in EriII and AndXXV provide constraints on 
the host halos. Since a split in multiple stellar subpopulations 
is not available for these two systems, the only constraint posed by present data 
is on the total mass within the half-light radius \citep{MW09,JW10,NA11,AA14}.
I use the mass estimator proposed by \citet{Ca16} (their eqn.~(17) and Table~3).
With a form similar to the estimator proposed by \citet{NA11,NA12},
this was tested on cosmological simulations and it appears to minimize bias and scatter.
Taking into account both 
observational and systematic uncertainties, the constraints on the total mass are 
 \begin{equation}
7.0 < \log \left[ M_{tot}(<1.77 R_h)/M_\odot\ \right] < 7.6 
 \label{massE}
 \end{equation}
for EriII, and 
 \begin{equation}
6.2 < \log \left[ M_{tot}(<1.77 R_h)/M_\odot \right] < 7.3 \ ,
 \label{massA}
 \end{equation}
for AndXXV, where $R_h$ are the galaxies' half light radii. 
I assume the halo has a parametric form 
\begin{equation}
\rho(r)={\rho_s\over{{\left({r\over r_s}\right)^{\alpha}}\left(1+{r\over r_s}\right)^{3-\alpha}}}\ ,
\label{density}
\end{equation}
in which the classical Navarro-Frenk-White density profile 
\citep[NFW, ][]{NFW96} corresponds to the case $\alpha=1$. For such cuspy haloes,
the 1-sigma regions shown {in} panel $a$ of Figure~1 illustrate the constraints~(\ref{massE},\ref{massA}). 
Here $\rho_s$ and $r_s$ are respectively the characteristic density and 
scale radius of the density profile, as in eqn.~(\ref{density}).
Full black dots in the same panel represent mean cosmological haloes at $z=0$, satisfying the
mass-concentration relation of cold DM haloes \citep[as compiled by ][]{AL16}. Points 
range between a virial mass $M_{200}$ of $10^7$ and $10^{11}~M_\odot$, in steps of 0.5 dex.
Using the same steps, smaller points illustrate the range allowed by
the scatter in the mass-concentration relation, for the same set of masses. 

EriII appears most compatible with a cold DM halo of $\log M_{200}/M_\odot=9$, 
having
\begin{equation}
\left\{
\begin{array}{rcr}
r_s~ {\rm kpc}^{-1}& = &1.58 \\
\log \rho_s~ M_\odot^{-1} {\rm kpc}^3& = & 7.07
\end{array}
\right. \ .
\label{NFWvals}
\end{equation}
This value of the virial mass is in good agreement with what would be inferred based on the dwarf's luminosity 
using abundance matching \citep[e.g.,][]{GK17,PJ16}.
Haloes with different inner slopes, $\alpha<1$, and compatible with the same mass constraint can be obtained as follows. 
The scale radius $r_s$ is kept fixed as in~(\ref{NFWvals}), as it would be if
the central density cusp is {removed} by feedback or scoured by the orbital evolution of gaseous 
massive clumps \citep[e.g.,][]{EZ01,MB04,Co11,NB15,DP16}. The characteristic density $\rho_s$ is adjusted as a function
of $\alpha$, so that the enclosed mass $M_{tot}(<1.77 R_h)$ remains constant. 

As shown by the blue lines in panel $a$ of Fig.~1, the low value of the velocity dispersion of AndXXV 
would suggest an unexpectedly low virial mass.
\citet{MC13} have already identified the peculiar properties of AndXXV, which is an outlier in the population
of Local Group dwarfs. They concluded that AndXXV is likely to have been recently affected by tides \citep[see also][]{MC14}. 
Indeed, the value of the virial mass obtained above assuming dynamical equilibrium 
appears exceedingly low. This is especially true when compared to the dwarf's brightness, which would instead suggest a
halo at least as massive as the one in EriII.

\subsection{Instantaneous tidal radii}

For a cluster with mass $M_c$ orbiting within the spherically symmetric potential $\Phi(r)$ and 
instantaneously at a galactocentric distance $r$, the nominal tidal radius is given by \citep[see e.g.,][]{FR11,NA15}
\begin{equation}
r_t = \left( {{G M_c}\over{\Omega^2(r)-d^2\Phi/dr^2}}\right)\ ,
\label{tidr}
\end{equation}
where $\Omega^2(r)=r^{-1}d\Phi /dr$. For a Keplerian gravitational potential generated by a mass $M$, 
Eqn~(\ref{tidr}) returns the classical $r_t=r(M_c/3M)^{1/3}$. However, the shape of the density profile 
should also be taken into account.

Panels $b$ and $c$ of Fig.~1 display contours
for the instantaneous tidal radius~(\ref{tidr}), measured in pc, in the plane of orbital distance $r$ versus slope of the 
density profile $\alpha$, as from eqn.~(\ref{density}). Following the analysis of Sect.~3.1, for EriII
I have assumed that, when $\alpha=1$, the halo has the properties of a mean cold DM 
halo with $\log M_{200}/M_\odot=9$, as in eqn.~(\ref{NFWvals}).  For different values of
$\alpha$, the dimensional scales are obtained as described in Sect.~3.1,
i.e. so to satisfy the kinematic constraint~(\ref{massE}). 
For AndXXV, the arguments in Sect.~3.1 are inconclusive. Given the dwarf's luminosity, 
the value $\log M_{200}/M_\odot=9$ is likely a lower bound to the original virial mass of the system,
and therefore represents a conservative choice with respect to the strength of the tidal field.
For this reason, panel $c$ of Fig.~1 also adopts $\log M_{200}/M_\odot=9$.
As to the clusters themselves, Fig.~1 assumes that the cluster in EriII has a mass of $M_c=5\times 10^3 M_\odot$ and the one in 
AndXXV has $M_c=25\times 10^3 M_\odot$, corresponding to a mass to light ratio of 
$M/L_V\approx2.5~M_\odot/L_\odot$.

\begin{figure*}
\centering
\includegraphics[width=\textwidth]{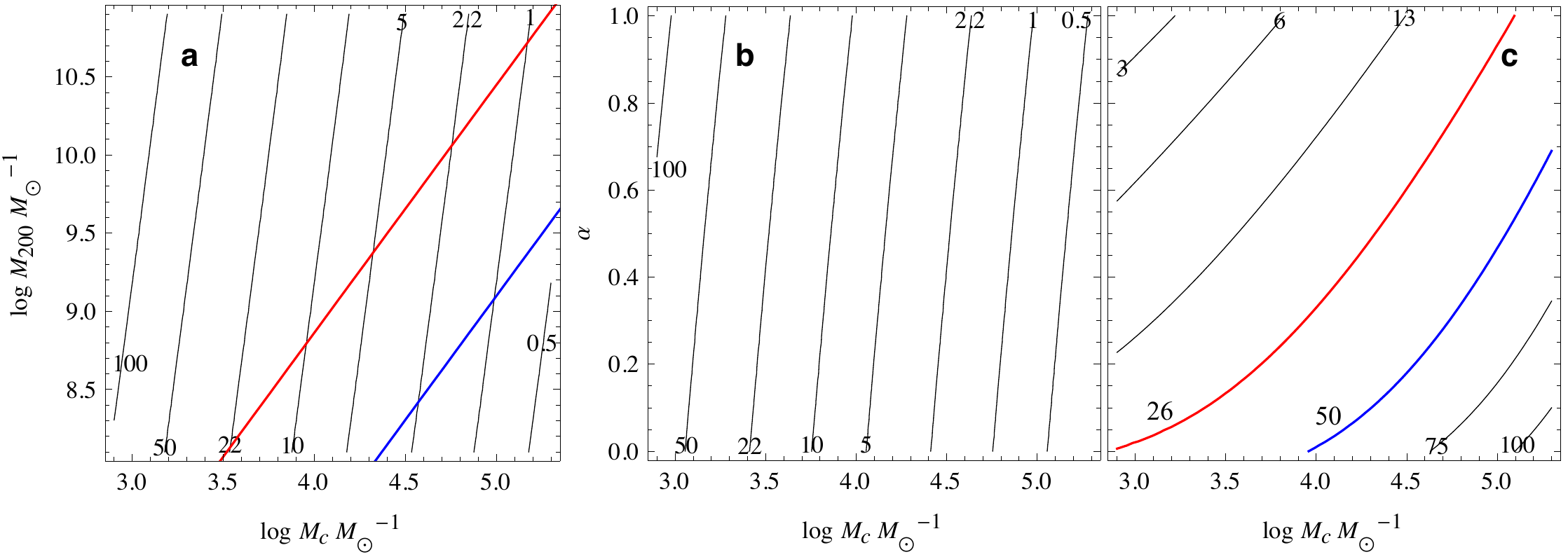}
\caption{Estimates of the dynamical friction timescale. Panel $a$ shows contours of $t_{fr}$ in Gyr,
i.e. the time it takes for a cluster with mass $M_c$ to sink from a radius of 250~pc to the center of mean cosmological 
NFW halo with virial mass $M_200$. The red and blue lines show where $r_t(r=250~{\rm pc})=2R_{h,c}$
in the same plane, respectively for EriII and AndXXV. Panel $b$ displays contours of $t_{fr}$ in haloes with different 
central density slopes $\alpha$. Panel $c$ shows the `stalling radii' $r_{stall}$, where dynamical
friction is suppressed, or equivalently, the tidal radius at that location, $r_t(r_{stall})$.
Models above these lines are likely to experience significant 
tidal mass loss on their way to the center of the host.}
\end{figure*}

The red and blue lines in panels $b$ and $c$ of Fig.~1 display the contours
\begin{equation}
r_t = 2\ R_{h,c} \ ,
\label{rt=2}
\end{equation}
where $R_{h,c}$ are the observed projected half light radii of the two clusters. These
lines are approximate divides between configurations in which the 
cluster experiences substantial tidal loss, if $r_t\ll 2\ R_{h,c}$, 
and the opposite regime in which the cluster does not fill the Roche lobe and is safe against the tidal field, $r_t\gg 2\ R_{h,c}$.
Cuspy density profiles result in threatening tidal fields, with a sharp demarcation between 
profiles with $\alpha<0.2$ and $\alpha\gtrsim0.2$. In order to survive for a time comparable with its age
in a $\alpha\gtrsim0.2$ cusp, this simplified analysis suggests that the cluster in EriII should remain at 
galactocentric distances $r\gtrsim500$~pc. The one in AndXXV at distances $r\lesssim700$~pc.
Furthermore, in the presence 
of a density cusp and a non-circular orbit, the addition of tidal shocking 
at pericenter is likely to significantly facilitate the disruption of the cluster. This is especially true
since the dynamical time in the central regions of the haloes in object is short, $t_{dyn}\approx0.1$~Gyr,
resulting in repeated injections of energy into the cluster. As a consequence,
clusters are very likely
to be quickly destroyed if they happen to orbit at radii where $r_t\ll 2\ R_{h,c}$.
At face value, this result is at odds with the observation 
that {\it both} clusters have a projected distance $d\lesssim50$~pc from the center. 
I will return on this aspect in a quantitative manner in Sect.~5.1.

On the other hand, Fig.~1 suggests that if hosts are cored or have very shallow density slopes, $\alpha\lesssim0.2$,
both clusters are free to orbit at any galactocentric distance.
In fact, for $\alpha\lesssim0.2$, panels $a$ and $b$ in Fig.~1 show that the instantaneous tidal radius $r_t$ is a non monotonic
function of the orbital radius $r$: the very central regions are safer than when $r\approx R_h$. 
This follows from the sign inversion of the eigenvalues of the tidal tensor, which mark the transition 
to a compressive tidal field when in a constant density environment \citep[e.g.,][]{Cha42,FR11}.
This simple analysis suggests that, despite their low mass and large sizes, 
the clusters in EriII and AndXXV would be safe against the tides in haloes 
with a very shallow cusp or a large core. More in general, clusters even more fragile than those considered here
could survive indefinitely if, helped by dynamical friction, they can manage to cross the region $r\approx R_h$
and reach the sheltered inner core region.

\subsection{Dynamical friction}

It is well known that dynamical friction is an important ingredient in the evolution of 
GCs in dwarf galaxies \citep[e.g.,][]{Tre75,Oh00,JL01,SS06,Go06,Ha11,dB14,Br14}. The scope of this section is limited to providing
estimates for the sinking times tailored on the problem at hand. These estimates are useful to guide the identification of viable evolution 
scenarios for the clusters, to be explored numerically in Sect.~5.

The standard understanding of dynamical friction is crystallized in Chandrasekhar's analytic formula \citep{Cha43,BT08}:
\begin{equation}
{{d v_c}\over{d t}}= -4\pi G^2\ {{M_c \rho}\over{v_c^2}}\ \log\Lambda\ f_{v<v_c} \ ,
\label{Chandra}
\end{equation}
where $M_c$ is the cluster mass, $v_c$ is the norm of its velocity, $\rho$ is the background density,
$\log\Lambda$ is the usual Coulomb logarithm and $f_{v<v_c}$ is the fraction of the background 
density with velocities $v<v_c$. Here, I seek to estimate the dynamical 
friction timescale $t_{fr}$ of clusters with different masses in haloes with different density profiles.  
To this end, I consider the simplified scenario in which:
\begin{itemize}
\item{the cluster mass $M_c$ remains constant during sinking, i.e., the cluster is not contemporarily affected by tides;}
\item{the orbit of the cluster evolves loosing both energy and angular momentum but conserving its circularity
$j=1$, i.e. it remains a circular orbit;}
\item{variations in the factor $\log\Lambda f_{v<v_c}$ during the orbital evolution are secondary and can be neglected;}
\item{the cluster orbits in the central regions of the density 
profile~(\ref{density}), where $\rho\sim{\rho_s}\left(r/r_s\right)^{-\alpha}$.}
\end{itemize}
For the sake of clarity, $j$ is the orbital circularity $j\equiv J/J_{circ}(E)$, where $J$ and $E$ are the 
orbital angular momentum and the energy of the cluster, and $J_{circ}(E)$ is the angular 
momentum of a circular orbit with energy $E$. 
Under the model assumptions above, the dynamical friction timescale $t_{fr}$ scales as follows: 
\begin{equation}
\left\{
\begin{array}{rcl}
t_{fr} &\propto& G^{-1/2}\ {{\rho_s^{1/2} r_s^3}\ M_c^{-1}}\ \left({r_{i}/ r_s}\right)^{3-\alpha/2}\ g(\alpha)\\
g(\alpha) &=& (6-\alpha)^{-1}(4-\alpha)(3-\alpha)^{3/2}
\end{array}\right. \ .
\label{Chandrasol}
\end{equation}
This is the time it takes for dynamical friction to bring a massive object from the radius $r_i$ 
to $r=0$. I calibrate eqn.~(\ref{Chandrasol}) on the result of an N-body
simulation in which a massive particle, $\log M_c/M_\odot=4.7$, is put on a circular orbit
with $r_{circ,i}=250$~pc in a cold DM halo as in eqn.~(\ref{NFWvals}). This sinks in a time 
$t_{fr}\approx1.9$~Gyr. An analogous massive particle the same initial orbital energy,
but on a very eccentric orbit, $j=0.3$, sinks in a very similar time, $t_{fr}\approx1.85$~Gyr 
(see Sect.~4 for details on the numerical setup).

Panel $a$ in Fig.~2 shows the sinking time $t_{fr}$ in Gyr for clusters of 
mass $M_c$ in mean cold DM haloes ($\alpha=1$) with virial mass $M_{200}$. The starting radius $r_{circ,i}$
is kept fixed at a physical distance of $250$~pc. At fixed $M_c$, $t_{fr}$ increases 
with the virial mass of the host $M_{200}$: while $\rho(r)$ increases with $M_{200}$, 
so does the orbital velocity $v_{circ}(r)$, at the denominator in eqn~(\ref{Chandra}). 
The dependence on $M_c$ is instead more marked: while 
dynamical friction can be ignored for clusters with $M_c\lesssim10^{3.7}~M_\odot$, while 
massive clusters with $M_c\approx10^5~M_\odot$ 
sink to the center very quickly. Clusters with masses $M_c\approx10^{4.5}~M_\odot$ would also sink towards
the center in a fraction of the Hubble time. However, as the numerical 
explorations of Sect.~5 show, the interplay between dynamical friction and
mass loss is significant in this regime and the estimates obtained here are lower limits for $t_{fr}$.
As in Fig.~1, the red and blue lines in panel 
$a$ of Fig.~2 display the curves $r_t(r)=2 R_{h,c}$, 
here assuming $r=r_{circ,i}=250$~pc. Models close and above these lines 
are unlikely to survive all the way to the center of haloes with $\alpha=1$.  

 Panel $b$ explores the dependence of $t_{fr}$ with $\alpha$. Similarly to panels $b$ and $c$ of Fig.~1,
 this uses the parameters~(\ref{NFWvals}) for $\alpha=1$, and adjusts $\rho_s(\alpha)$ so to satisfy the mass 
 constraint~(\ref{massE}) for $\alpha<1$. The dependence of $t_{fr}$ on $\alpha$ is rather mild.
 This is again due to the competing dependences on the local density $\rho(r)$ and local circular velocity $v_{circ}(r)$ 
 in eqn~(\ref{Chandra}), which both decrease with $\alpha$. 
 
 Depending on the cluster mass and on the details of the host density profile, 
 dynamical friction can only bring the cluster to a finite distance from the center, $r_{stall}$, where it becomes
 strongly suppressed and the sinking process `stalls'. This behavior is due to the formation of a central core 
 in the host density profile, on the scale of $r_{stall}$, as a consequence of the energy and angular momentum transfer from the 
 cluster itself \citep[e.g.,][]{EZ01,Go10}. In shallow density profiles, dynamical friction is suppressed 
 by the emergence of orbital resonances \citep[e.g.,][]{HG98,JR06,In09,Co12,Pe16}, which effectively
 halt the sinking process. This behavior is not captured by the Chandrasekhar approximation~(\ref{Chandra}),
 and therefore not taken into account in the estimates above.
 Analytical arguments and numerical studies \citep[see e.g.,][]{Go10,Pe16} show that `core stalling' 
 happens at the radius $r_{stall}$ where
\begin{equation}
r_t(r_{stall})\approx r_{stall} \ .
\label{rmin}
\end{equation}
Panel $c$ shows the stalling radii predicted by this equation, in pc, as a function of $M_c$ and $\alpha$.
By definition, these are also the values of $r_t(r_{stall})$. The red and blue lines show the contours $r_t(r_{stall})=2R_{h,c}$ 
for the clusters in EriII and AndXXV. Models above these lines are likely to experience significant 
tidal mass loss on their way to the center of the host. 

\section{Numerical setup}

In this Section, I describe the setup used for the N-body experiments presented in this paper. 
All simulations are collisionless N-body simulations, executed with Gadget-2 \citep{VS05}.
Each features a spherically symmetric host and a spherically symmetric star cluster.

Initial conditions for the star cluster are
generated assuming an isotropic Plummer phase space distribution function \citep[e.g.,][]{Pl11,HD87},
for a density profile
\begin{equation}
\rho_c(r)={{3 M_c}\over{4 \pi r_c^3}}\left(1+{r^2\over{r_c^2}}\right)^{-5/2} \ .
\label{plummer}
\end{equation}
The total mass $M_c$ and the core size $r_c$ are free parameters. The projected half light radius 
for a Plummer model is $R_h=r_c$, while the 
spherical half light radius is $r_h\approx1.31 r_c$. In the runs described in Sect.~3.3 and used to calibrate the dynamical
friction timescale~(\ref{Chandrasol}), the star cluster is represented by a single massive particle. 

The difficulty presented by these simulations is that both cluster and host 
should be resolved with live particles, and that the individual masses of these particles 
should be comparable. This makes for a significant computational challenge,
by bringing the total number of particles to $N_{tot}\sim10^9$.
A live cluster is necessary to capture tidal stripping; a live halo for dynamical friction. 
In a static background potential, dynamical friction should be included by hand. 
However, a genuine accounting of the sinking process is to be preferred here, as 
the interplay between sinking and mass loss decides the fate of the cluster (see Sect.~5).
Particles in host and cluster should have comparable masses to avoid 
artificial dynamical heating of the cluster \citep[see e.g.,][]{TB16,KL17}, 
which would substantially facilitate its disruption. 

{I} circumvent this computational issue by combining the 
use of a static potential and live particles for the host halo. The clusters inhabit the
central regions of the halo, therefore resolving the host's internal dynamics at large radii
is unnecessary. 
I mimic the technique of particle tagging \citep[often adopted in studying 
the assembly of the stellar halo of galaxies, e.g.,][]{JB01,AC10,AC13,NA17} and order 
the host's mass by binding energy. Only a fraction $f_{live}$ of most bound mass 
is resolved in live particles, the remainder of the host mass is replaced by a static background potential. 
As this selection in based on an integral of the motion, the live particles are in equilibrium 
within the combination of their own potential and of the associated static background.
The latter contributes a force
\begin{equation}
F_{bkg}(r) = F_{tot}(r) \left[1-{{M_{live}(< r)}\over {M_{tot}(<r)}}\right] \ .
\label{force}
\end{equation}
where $F_{tot}$ is the total force and $M_{tot}(<r)$ is the total enclosed mass.
Here, `total' refers to the complete target density profile, while $M_{live}(<r)$ is the enclosed mass in
live particles alone.

\begin{figure}
\centering
\includegraphics[width=.8\columnwidth]{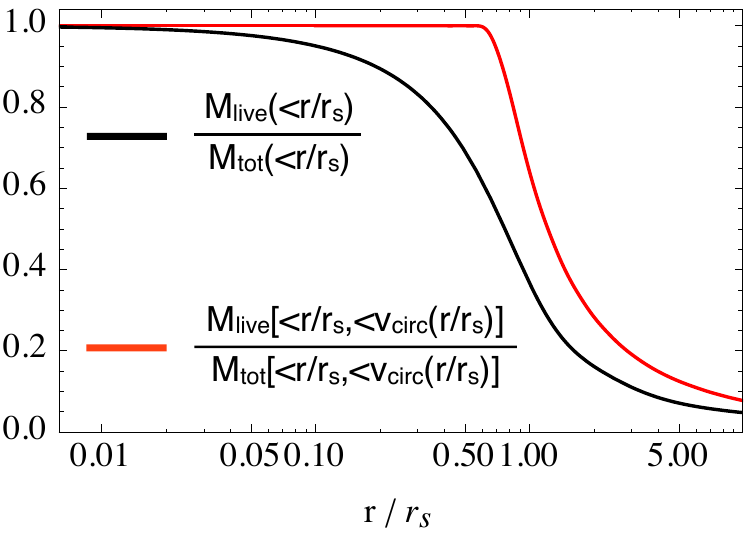}
\caption{Black line: the fraction of enclosed mass in live particles for 
an NFW halo and a fraction $f_{live}=3\%$. For the same halo and $f_{live}$,
the red line shows the fraction of enclosed mass resolved in live particles with $v<v_{circ}(r)$
(see text for more details).
Where this fraction is $\approx 1$, dynamical friction is accurately described.}
\end{figure}

The minimum necessary fraction $f_{live}$ is fixed by the requirement that the mass responsible for 
dynamical friction in the regions of interest should be resolved in live particles.
As captured by eqn~(\ref{Chandra}), this mass is represented by the `slow' fraction of the local density:
$\rho(r) f_{v< v_c}$. Assuming $\alpha=1$, the red line in Fig.~3
quantifies the fraction of slow particles that are retained as live for $f_{live}=3\%$. 
In more detail, this is the ratio between 
\begin{itemize}
\item{$M_{live}\left[<r/r_s,v<v_{circ}(r/r_s)\right]$, i.e. the mass of the host (i) resolved in live particles; 
(ii) enclosed in the radius $r/r_s$; (iii) having orbital velocities smaller than the circular velocity at the enclosing radius, $v_{circ}(r/r_s)$.}
\item{$M_{tot}\left[<r/r_s,v<v_{circ}(r/r_s)\right]$: the total mass of the host satisfying (ii) and (iii).}
\end{itemize}
A fraction $f_{live}=3\%$ is sufficient to fully describe dynamical 
friction at radii $r/r_s\lesssim0.6$. This can be compared with the ratios $R_h/r_s$,
$\approx 0.18$ for EriII and $\approx 0.4$ for AndXXV, ensuring that a choice of
$f_{live}=3\%$ is appropriate. For this value, 
the black line in Fig.~3 shows the fraction $M_{live}(<r/r_s)/M_{tot}(<r/r_s)$, appearing in the equation for the
background force component~(\ref{force}). 
Initial conditions for the live particles are 
generated using the phase space distribution function of the target density profile, 
which is obtained under the assumption of orbital isotropy through the standard Eddington 
inversion \citep[e.g.,][]{Edd16,LW00}. This strategy entirely eliminates the computational issue 
described above. At the same time it allows me to genuinely capture the effects of dynamical friction.

\section{Cuspy haloes}

In this Section I assume that EriII and AndXXV have cuspy haloes, 
and numerically explore consequent evolution scenarios for the star clusters.

\begin{figure*}
\centering
\includegraphics[width=\textwidth]{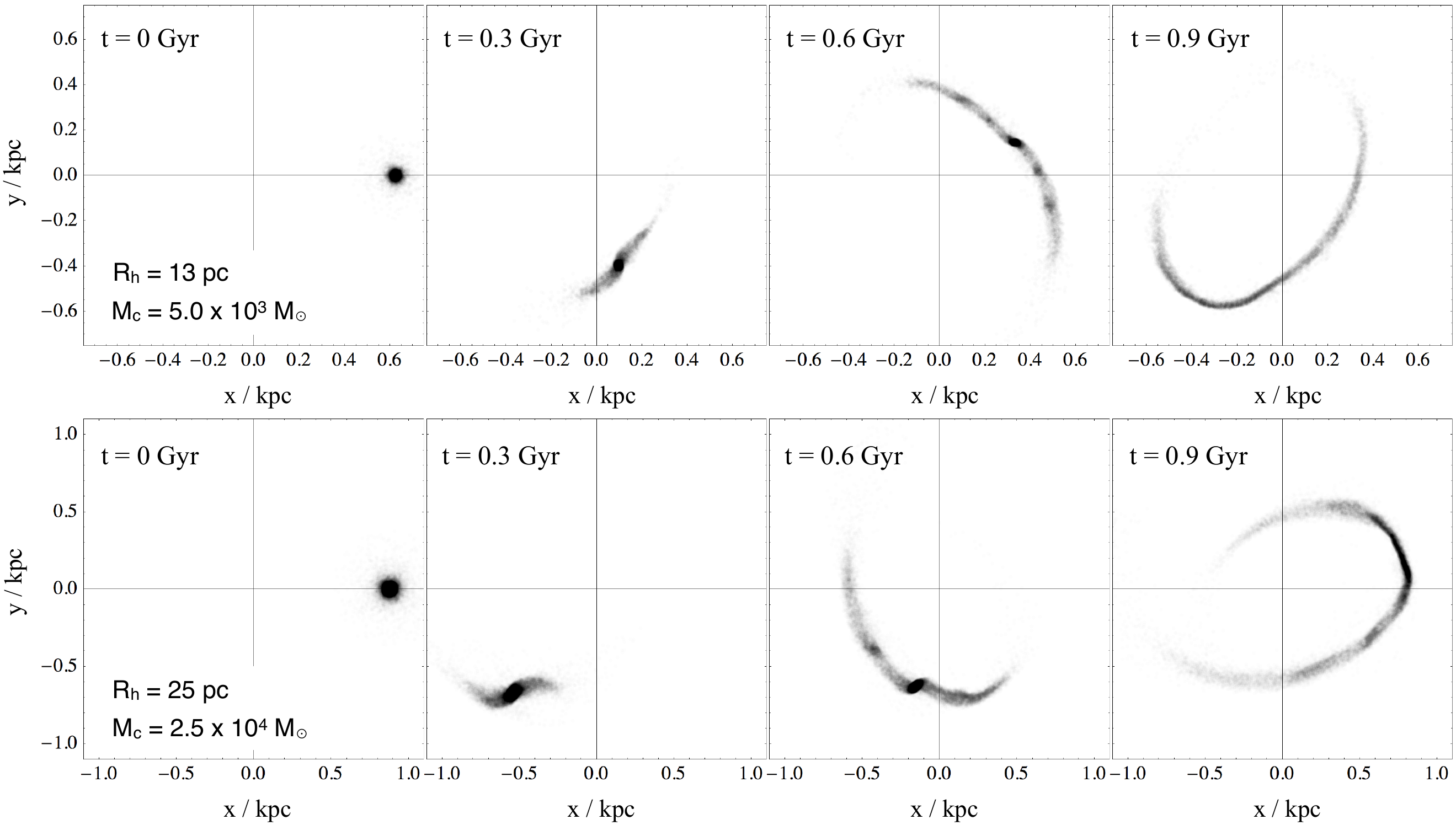}
\caption{Examples of quick tidal disruption for star clusters as observed in EriII (upper panels) 
and in AndXXV (lower panels) in NFW haloes.
The cluster are put on high circularity orbits ($j=0.9$) with energies $r_{circ}(E)=0.5$~kpc for EriII and 
$0.7$~kpc for AndXXV. Both clusters are fully disrupted by tides within 1 Gyr.}
\end{figure*}

\subsection{The clusters formed as they are}

I first consider the case in which the clusters formed with properties
similar to those currently observed: $\log M_c/M_\odot= 3.7$ and
$R_{h,c}=13$~pc in EriII, $\log M_c/M_\odot= 4.4$ and
$R_{h,c}=25$~pc in AndXXV. 
Fig.~4 shows what would happen if these clusters are put in orbit 
within a host halo with $\alpha=1$ and parameters like in eqn~(\ref{NFWvals}). 
Top panels refer to EriII, bottom panels 
to AndXXV. In both, the orbit has a very high circularity,
$j=0.9$, which keeps the cluster away from the central regions. 
Orbital energies are motivated from the analysis of Sect.~3.1 and
and correspond to $r_{circ}(E)=0.5$~kpc for EriII and $r_{circ}(E)=0.7$~kpc for AndXXV.
Here $r_{circ}(E)$ is the radius of a circular orbit with energy $E$, and the chosen values 
satisfy $r_t\approx2R_{h,c}$.

Fig.~4 confirms the findings of Fig.~1: a cuspy halo represents an unavoidable threat
for star clusters that are as faint and as extended as in EriII and AndXXV.
Both clusters disrupt completely within a Gyr. 
The haloes in Figure~4 have $\alpha=1$. However, Fig.~1 shows that the location of the contour $r_t\approx2R_{h,c}$
is insensitive to the density slope for $\alpha\gtrsim 0.2$. For them to survive 
for longer times in cuspy haloes the clusters should {\it constantly} orbit at larger radii, 
with higher orbital energies.
This result shows that the observed locations of the clusters in EriII and AndXXV 
are due to projection effects in this scenario. 

What is the likelihood of this?
In other words, what is the the probability to observe the clusters so close to the center if they are 
forced to orbit at $r > r_{min}$? A rough estimate can be obtained by considering the phase 
space of a system that has been fully voided within the loss cone corresponding to $r_{peri}<r_{min}$:
\begin{equation}
g(E, J)=\left\{
\begin{array}{lc}
0 & r_{peri}(E,J)<r_{min} \\
f(E, J) & r_{peri}(E,J)\geq r_{min}
\end{array}
\right. \ .
\label{forbit}
\end{equation}
Orbits that bring the cluster to a pericenter $r_{peri} < r_{min}$ are not 
viable as they result in quick disruption. 

I take $f$ so that it describes the galaxy's stars: when integrated over the entire phase space
$f$ generates an approximately Plummer density profile with the correct half-light radius. In particular, 
I take $f$ to have an isotropic lowered isothermal form, which has been shown to provide a good 
description of both density and kinematic profiles of dSphs \citep[e.g.,][]{NA11,NA12}.
I take $r_{min}=0.5$~kpc for EriII and and $r_{min}=0.7$~kpc 
for AndXXV and assume that all orbits with $r_{peri}\geq r_{min}$ are equally viable.
This is equivalent to requiring that the lifetime of the clusters is $\gtrsim$1~Gyr.

The fraction of mass the phase space density $g$ generates within a projected distance $\delta_{LOS}$ from the center
represents the probability of observing the cluster at an instantaneous projected location
$d<\delta_{LOS}$. Such probability is shown in the top panel of Fig.~5, in red for EriII and in blue for AndXXV. 
The yellow shaded area shows the range selected by the observations: $\delta_{LOS}<60$~pc.
For either systems individually, the probability of observing the cluster at such small radii is very small: $p\lesssim 0.4$\%.
As the two systems are independent, the probability of observing both this close to the center is entirely negligible. 
This excludes the possibility that the clusters in EriII and AndXXV formed in cuspy haloes with structural 
properties similar to those currently observed and have a lifetime $\gtrsim1$~Gyr. 
A possible way to escape this is that the clusters formed at the center of their respective host haloes, 
which is examined in Sect.~5.1.1.

It is interesting to consider kinematic predictions of the different scenarios, providing means to test them.
This is especially the case for EriII, since the dwarf's mass is substantially more secure than in the case of 
AndXXV, as discussed in Sect.~3.1.  
The red histogram in the bottom panel of Fig.~5 shows the probability distribution of the cluster's LOS velocity,
relative to the systemic velocity of EriII $v_{LOS, sys}$,
\begin{equation}
\delta v_{LOS}  \equiv v_{LOS} - v_{LOS, sys} \ ,
\label{dvlos}
\end{equation}
as obtained from the phase space distribution function $g$, for $d<60$~pc.
In gray, the same probability distribution is shown for all EriII stars at $d<60$~pc, as generated by $f$.
The distributions are similar, but large LOS velocities are somewhat disfavored for the cluster.
In this scenario, the internal velocity dispersion of the clusters is as implied 
by their stellar mass. For a mass to light ratio $M/L_V=2.5~ M_\odot/L_\odot$, 
the internal velocity dispersions are 
\begin{equation}
\sigma_c\approx\left\{
\begin{array}{lc}
0.53~{\rm kms}^{-1} & {\rm in~EriII}\\
0.85~{\rm kms}^{-1} & {\rm in~AndXXV}
\end{array}
\right. \ .
\label{sigmac}
\end{equation}

Finally, in this scenario, it is fair to ask whether DM subhaloes could represent an additional threat 
for the clusters. If I assume the population of bound substructures in cold DM haloes with $\log M_{200}/M_\odot=9$ 
is a scaled version of the one in Milky Way sized haloes \citep{JD08,VS08}, close encounters that could 
seriously disturb the clusters are extremely rare. This is due to the combination of the high relative velocities and 
of the preferentially higher orbital energies of subhaloes, which are rare in the central regions \citep{VS08}.

\begin{figure}
\centering
\includegraphics[width=.85\columnwidth]{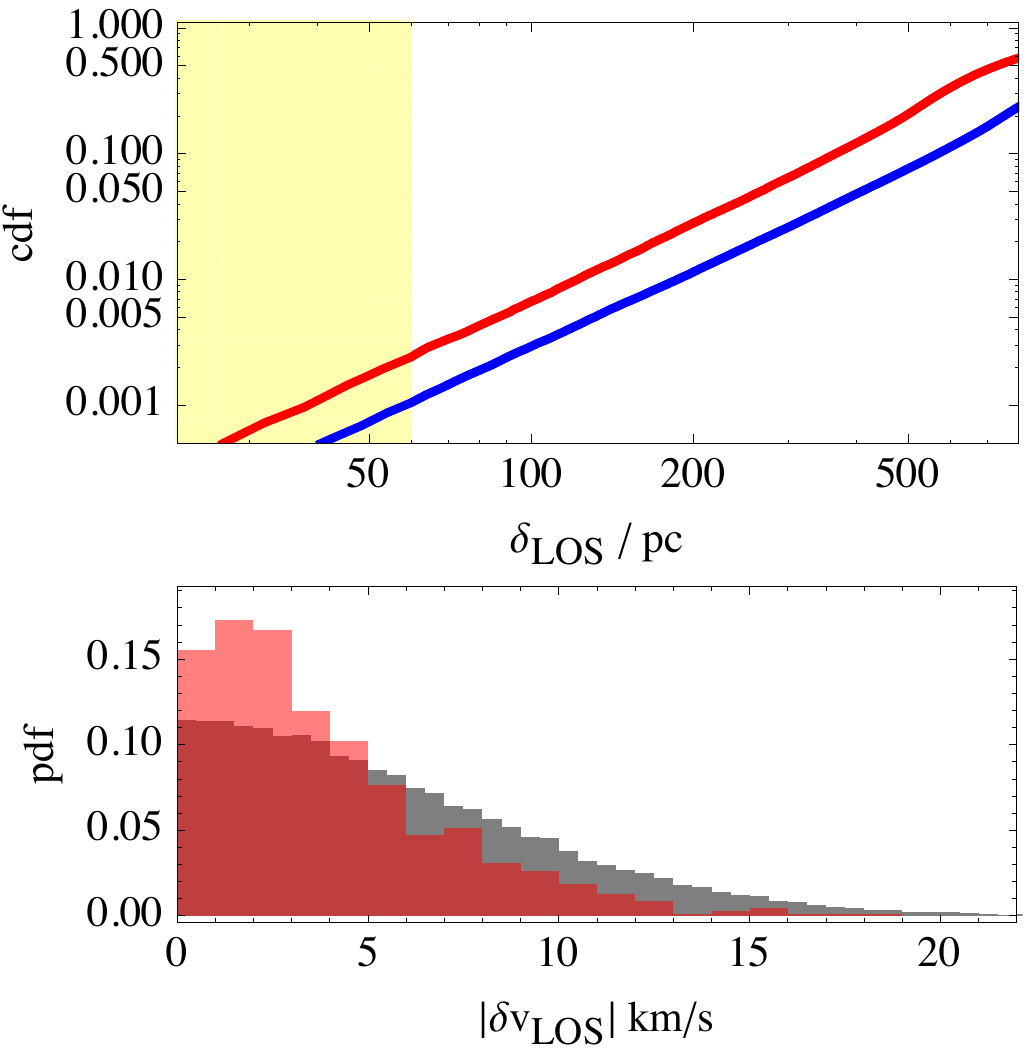}
\caption{Top panel: in red (blue), the probability of observing the cluster in EriII (in AndXXV)
at a projected distance $d<\delta_{LOS}$ from the center of a cuspy halo ($\alpha\gtrsim0.2$, virial mass $M_{200}\approx 10^9 M_\odot$), if its lifetime is $\gtrsim$1~Gyr. Bottom panel: in red, the prediction for the LOS velocity of the cluster
in EriII, compared to the LOS velocity of stars at comparable distances from the center, in gray.}
\end{figure}
\begin{figure}
\centering
\includegraphics[width=.85\columnwidth]{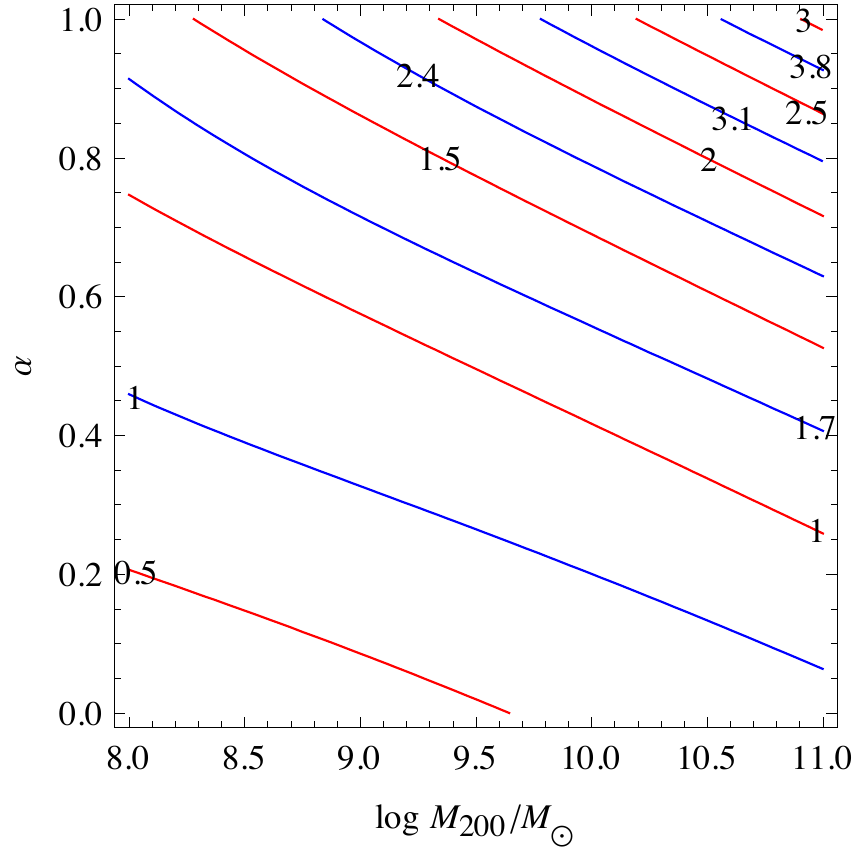}
\caption{Predicted values for the internal velocity dispersion of the clusters, in kms$^{-1}$,
if these formed at rest at the center of the halo. Red lines refer to the case of the cluster in EriII and blue lines to the 
case of AndXXV. The displayed values assume the clusters are dark matter dominated. For $M/L=2.5 M_\odot/L_\odot$
and no dark matter, the predicted values of the internal velocity dispersions are as in eqn.~(\ref{sigmac}).  }
\end{figure}

\begin{table*}
\renewcommand{\thetable}{\arabic{table}}
\centering
\caption{Suite of simulations with live, $\alpha=1$ halos. The table lists the initial properties of the clusters and 
the final fate of the cluster, where $D$ corresponds to tidal disruption, while $S$ is a surviving cluster.
All surviving cluster manage to sink to the center of the halo.} \label{suiteNFW}
\begin{tabular}{cc}
\tablewidth{0pt}
\hline
\hline
$\left( \log M_{c,i}/M_\odot, j_{c,i}, r_{h,c,i}/{\rm pc}\right)$  & outcome\\
\hline
$(\left\{3.9, 4.1\right\}, 1.0, 10.0)$ & $\left\{S, S\right\}$\\
$(4.3, \left\{0.3, 0.5, 0.7, 0.8, 0.9, 0.95, 1.0\right\}, 10)$ & $\left\{D, D, D, D, D, D, D, S, S\right\}$\\
$(\left\{4.5, 4.55, 4.575, 4.577, 4.579, 4.582, 4.588, 4.6\right\}, 0.3, 10)$ & $\left\{D, D, D, S, S, S, S, S\right\}$\\
$(4.7, 0.3, \left\{10, 12, 12.8, 13.5, 15\right\})$ & $\left\{S, S, S, D, D\right\}$\\
$(\left\{4.7, 4.8\right\}, 0.9, 20.0)$ & $\left\{D, S\right\}$\\
$(4.8, 0.9, 22.5)$ & D\\
$(4.8, 1.0, 25.0)$ & D\\
$(4.9, 0.3, \left\{15, 20\right\})$ & $\left\{S, D\right\}$\\
$(4.9, 0.6, \left\{15, 20, 22.5\right\})$ & $\left\{S, S, D\right\}$\\
$(5.0, 0.3, \left\{10, 20, 21, 22, 23, 25\right\})$ & $\left\{S, S, D, D, D, D\right\}$\\
$(5.0, \left\{0.8, 0.85, 0.9\right\}, 25)$ & $\left\{D, S, S\right\}$\\
$(5.0, 1.0,  \left\{10, 25, 30, 35\right\})$ & $\left\{S, S, S, D\right\}$\\
$(5.3, 0.3,  \left\{25, 30, 35, 40\right\})$ & $\left\{S, S, S, D\right\}$\\
$(5.3, 0.6,  \left\{30, 35\right\})$ & $\left\{S, S\right\}$\\
\hline
\hline
\end{tabular}
\end{table*}

\subsubsection{The clusters formed as they are, at the center }

If sitting at rest at the bottom of the potential, the cluster would not be subject to tides.
The clusters could therefore have formed at the center of the cusp, and this would allow 
them to survive indefinitely, despite their low densities. 
By definition, in this scenario 
\begin{equation}
\left\{
\begin{array}{lcr}
d & = & 0\\
\delta v_{LOS} & = & 0
\end{array}
\right. \ .
\label{rest}
\end{equation}
The predicted negligible offset is somewhat ad odds with observations,
but it is possible the measured values are spurious, caused for example by
lopsided stellar distributions.

Contrary to the scenario of Sect.~5.1, in this case the cluster could show inflated dynamical mass to light ratios,
due to the high central dark matter densities. 
Fig.~6 shows shows the predicted values of the cluster's internal velocity dispersion, in kms$^{-1}$,
in the plane of the halo's virial mass and central slope. Halos are constructed as described 
in Sect.~3.1: haloes of the same virial mass have the same total mass within the galaxy's half
light radius.  Fig.~6 assumes that the process of cluster formation does not alter the properties of the central cusp, 
and ignores the stellar content of the clusters. Therefore, when the displayed values are higher than those
implied by stars alone in eqn~(\ref{sigmac}), the clusters are DM dominated.
As in previous Figures, red refers to EriII and blue to AndXXV. 
For virial masses as selected by the constraint~(\ref{massE}),
$M_{200}\approx 10^9 M_\odot$, most cusps $\alpha\gtrsim 0.4$ result in dark matter dominated
clusters, and the effect is especially significant in the case $\alpha=1$, offering means
to test this observationally.

\begin{figure*}
\centering
\includegraphics[width=.8\textwidth]{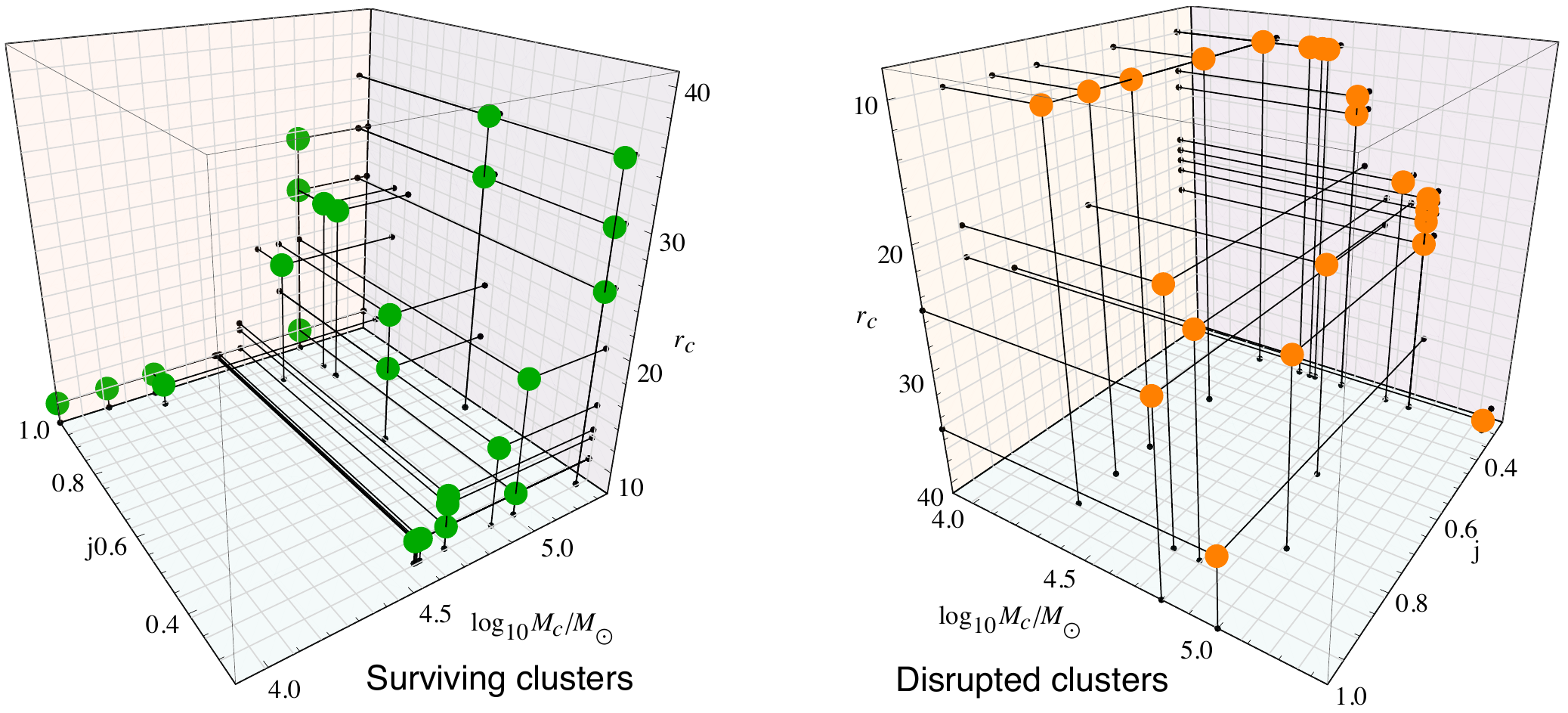}
\caption{A three dimensional view of the initial conditions that result in surviving clusters, left panel, against those that
lead to its tidal disruption, right panel.}
\end{figure*}
\begin{figure*}
\centering
\includegraphics[width=\textwidth]{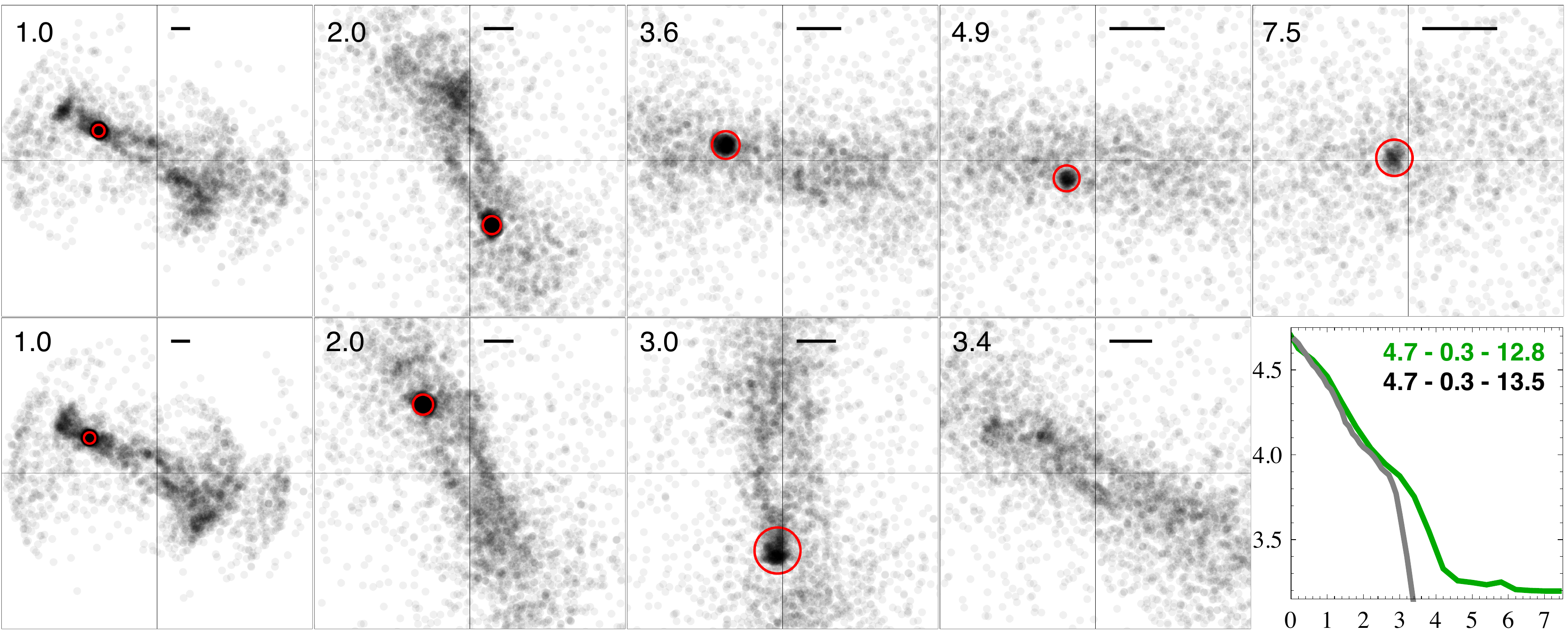}
\caption{Snapshots from the evolution of two clusters sharing initial mass, $M_{c,i}=10^{4.7}M_\odot$, and initial orbital circularity,
$j=0.3$, but having different initial sizes: $r_{c,i}=12.8$~pc in the top panels and $r_{c,i}=13.5$~pc
in the bottom panels. The time in Gyr of each snapshot is displayed in the upper-right corner of each panel. 
Black horizontal bars display a length of 50 pc. Red circles identify the remnants and have a radius of $2 R_{h,c}$
at that time. The bottom-right plot shows the full time-evolution of the mass of the two clusters (time, on the horizontal axis 
is in units of Gyr; mass, on the vertical axis, is in units of $\log M_{c}/M_\odot$). }
\end{figure*}

\subsection{The clusters were originally more massive}

In this Section I explore the possibility that the clusters had different properties at formation: i.e.
that their mass was initially higher and/or their half light radius initially smaller. Both these possibilities 
would help them survive for longer times in a cuspy halo. 
As shown in Fig.~2, for high enough masses the dynamical friction timescale $t_{fr}$ becomes 
substantially shorter than a Hubble time: it is in theory possible that the clusters might have sunk to the center of the host 
while losing mass because of the tides. Here, I explore on whether this scenario can actually reproduce
at the same time both the low mass and the large sizes of the clusters in EriII and AndXXV.

Motivated by the constraint~(\ref{massE}), I concentrate on halos with parameters like in eqn~(\ref{NFWvals}),
corresponding to $M_{200}\approx 10^9 M_\odot$, and take the case of the classical NFW cusp, $\alpha=1$. 
The combination of Figs.~1 and~2 would suggest that results obtained in this way might be at least 
qualitatively similar to the more general case of $\alpha\gtrsim0.2$. 
I fix the initial orbital energy of the cluster at $r_{circ}(E)=0.25$~kpc. In other words, the simulations of this Section
investigate the fate of the cluster if this, while losing energy, approaches the threshold $r_{circ}(E)=0.25$~kpc with a given set of properties.
Once these quantities are fixed, I am left with a three dimensional parameter space, 
featuring the initial cluster mass $M_{c,i}$, the initial orbital circularity $j_{c,i}$ and the initial three-dimensional half light radius $r_{h,c,i}$.
All runs feature a star cluster with $N_c=5\times10^3$ or  $N_c=10^4$ equal mass particles. 
The host has $N_h=10^6$ equal mass particles, representing the most bound 3\% of its mass, as 
described in Sect.~4.

Table~1 lists the initial properties of all the runs explored. Runs that share 2 of the three 
free parameters $\left( \log M_{c,i}/M_\odot, j_{c,i}, r_{h,c,i}/{\rm pc}\right)$ are
listed on the same line using brackets. The `outcome' of the simulation summarizes the  
fate of the cluster: $D$ indicates a cluster that is disrupted by tides before 15 Gyr, $S$ refers to a cluster that
survives. In fact, in all explored cases surviving clusters have also sank to the center within 15 Gyr.
The outcome defines two separate regions in the three-dimensional parameter space,
and the set of explored initial conditions is aimed to best identify this demarcation.
Fig.~7 compares these two opposite volumes: the left panel shows with green points the simulations 
with outcome $S$, the right panel collects all runs with outcome $D$. 
In the left panel, black lines connect individual runs with outcome $S$ to all models
with higher initial mass $M_c$, higher initial circularity $j$ and smaller initial half light radii $r_c$. 
All these models would also survive. Similarly, individual runs with outcome $D$ are connected to all
models with lower initial mass $M_c$, lower initial circularity $j$ and larger initial half light radii $r_c$,
which are equally destined to tidal disruption.
\begin{figure*}
\centering
\includegraphics[width=\textwidth]{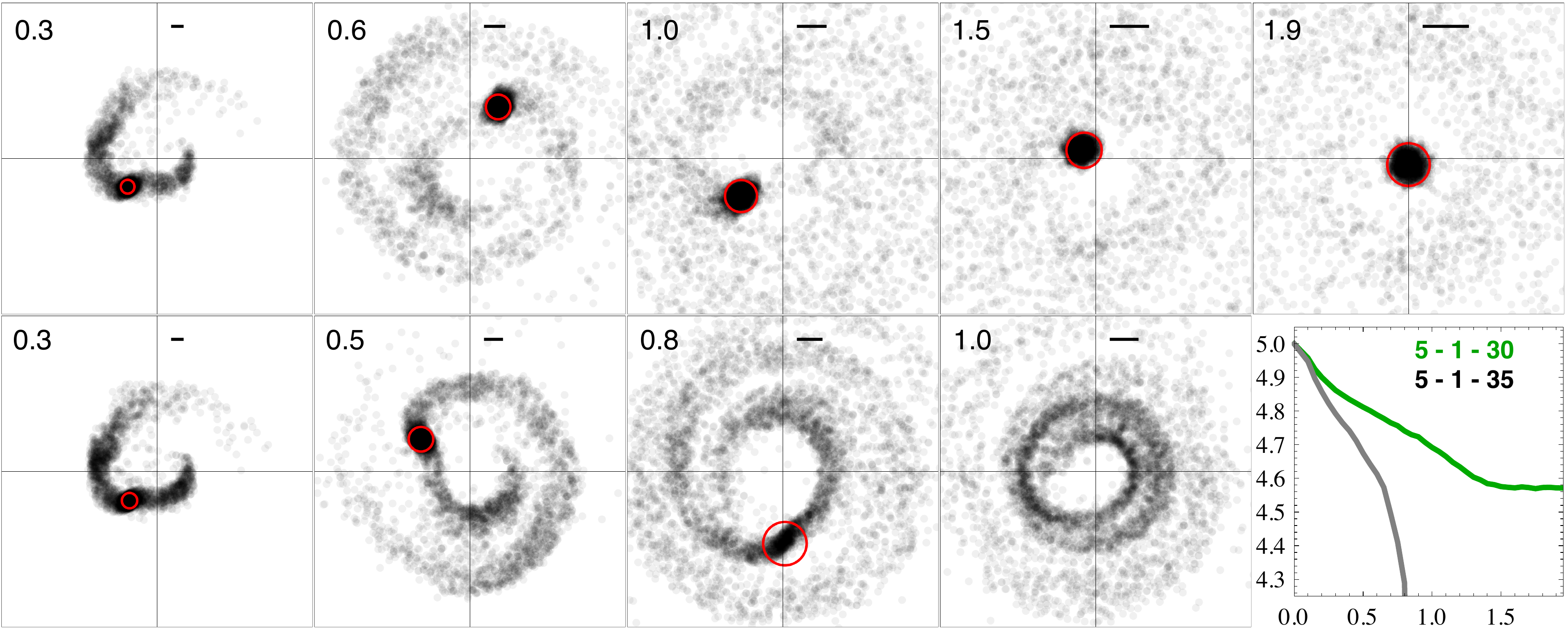}
\caption{Snapshots for the evolution of two clusters sharing initial mass, $M_{c,i}=10^{5}M_\odot$, and initial orbital circularity,
$j=1$, but having different initial sizes: $r_{c,i}=30$~pc in the top panels and $r_{c,i}=35$~pc in the bottom panels. Symbols and colors as in Fig.~8.}
\end{figure*}

Clusters with low mass, $M_{c,i}\lesssim10^{4.3}M_\odot$ only survive if concentrated, $r_{c,i}\lesssim10$~pc, and
initially on orbits that are essentially circular, $j\gtrsim0.95$. A circularity of only $j=0.9$ is capable of disrupting all
clusters with $M_{c,i}\leq10^{4.3}M_\odot$ and/or $r_{c,i}\geq10$~pc.
Clusters with intermediate initial mass, $10^{4.5} \lesssim M_{c,i}/M_\odot \lesssim10^{4.9}$, can manage to 
sink to the center even if initially on eccentric orbits, $j\geq0.3$, but need to be concentrated enough.
Dynamical friction and tidal mass loss are strongly coupled in this regime: a cluster that looses mass 
sinks more slowly, which in turn allows for further stripping before the halo center can be reached. 
As a consequence, small changes to the initial properties of the cluster result in different outcomes, 
and, when the cluster survives, in different final properties. An example of this is provided by Fig.~8, which illustrates the 
evolution of two clusters sharing an initial mass of $M_{c,i}=10^{4.7}M_\odot$ and an initial orbital circularity of
$j=0.3$, but having slightly different initial sizes: $r_{c,i}=12.8$~pc in the upper panels and $r_{c,i}=13.5$~pc
in the lower panels. Each panel displays a snapshot of the simulation, corresponding to the time indicated in the 
upper-left, in Gyr. The black bars display a length of 50 pc, while the red circles identify the remnants, 
and have a size of $2R_{h,c}$ at that time. No bound remnant is identified in the bottom panels at $t=3.4~$Gyr: 
the less concentrated cluster has been completely shredded by that time.
The bottom-right panel shows the mass evolution of both clusters: the abscissa indicates time in Gyr and 
masses are in units of $\log M_c/M_\odot$. While the concentrated cluster manages 
to sink to the center of the halo and survives thereafter with little to subsequent mass evolution, 
the small difference in the initial size causes the second cluster 
to be disrupted by tides.

High mass clusters, $M_{c,i}/M_\odot \gtrsim10^{5}$ are in general capable of sinking to the center,
unless they are exceedingly extended for their orbit. An initial half light radius of $r_{c,i}\leq20$~pc
allows clusters with $M_{c,i}\geq10^{5}M_\odot$ to survive, despite initial orbital eccentricities as low as $j\geq0.3$.
The boundary with models that lead to disruption is however thin: if $r_{c,i}\geq25$~pc, a cluster with the same 
mass would be disrupted unless initially on an orbit with very high circularity, $j\geq0.85$.
Figure~9 compares the cases of two clusters on initially circular orbits, $j=1$, 
which share an initial mass of $M_{c,i}=10^{5}M_\odot$,
but have different sizes: $r_{c,i}=30$~pc in the upper panels and $r_{c,i}=35$~pc in the lower panels.
The second cluster is exceedingly extended to survive, despite its high mass and the fine tuned, benign orbit.
Finally, clusters as massive as $M_{c,i}\geq10^{5.3}M_\odot$ manage to sink despite initial orbital eccentricities $j\geq0.3$,
as long as $r_{c,i}\lesssim35$~pc. 

I do not explore higher values of $M_{c,i}$, as these would make the cluster comparable or higher in mass 
than the total stellar population of the host galaxies. This limit is especially stringent for EriII, 
$L_V\approx 6\times10^4 L_\odot$: an initial mass $M_{c,i}\approx10^{5}M_\odot$
implies that the largest majority of the dwarf's stars once belonged to the cluster itself.

\begin{table*}
\renewcommand{\thetable}{\arabic{table}}
\centering
\caption{Suite of simulations with live, $\alpha=1$ halos resulting in surviving, sunk clusters. Column 1 lists the initial conditions;
Columns 2 and 3 list the mass and half light radius of the cluster once it has reached the halo center; Column 4 lists the time 
it takes for it to sink.} \label{suiteNFW}
\begin{tabular}{cccc}
\tablewidth{0pt}
\hline
\hline
$\left( \log M_{c,i}/M_\odot, j_{c,i}, r_{h,c,i}/{\rm pc}\right)$  & $\log M_{c,f}/M_\odot$ & $R_{h,c,f}/$pc & $t_{fr}/$Gyr\\
\hline
$(\left\{3.9, 4.1\right\}, 1.0, 10.0)$ & $\left\{3.4, 3.9\right\}$ &  $\left\{4.7, 6.1\right\}$& $\left\{14.9, 9.2\right\}$\\
$(4.3, \left\{0.95, 1.0\right\}, 10)$ & $\left\{4.0, 4.1\right\}$& $\left\{6.4, 6.8\right\}$& $\left\{6.4, 5.6\right\}$\\
$(\left\{4.577, 4.579, 4.582, 4.588, 4.6\right\}, 0.3, 10)$ & $\left\{3.8, 3.8, 3.9, 3.9, 3.9\right\}$& $\left\{6.1, 6.7, 6.7, 7.0, 7.0\right\}$& $\left\{9.4, 8.2, 4.9, 4.4, 4.1\right\}$\\
$(4.7, 0.3, \left\{10, 12, 12.8 \right\})$ & $\left\{4.4, 4.0, 3.2\right\}$ & $\left\{8.3, 8.1, 5.3\right\}$& $\left\{2.3, 2.8, 4.3\right\}$\\
$(4.8, 0.9, 20.0)$ & 4.4 & 10.0 &2.2\\
$(4.9, 0.3, 15)$ & 4.5& 9.3 & 1.6\\
$(4.9, 0.6, \left\{15, 20\right\})$ & $\left\{4.6, 4.4\right\}$& $\left\{9.6, 10.2\right\}$& $\left\{1.5, 1.8\right\}$\\
$(5.0, 0.3, \left\{10, 20\right\})$ & $\left\{4.9, 4.4\right\}$& $\left\{8.8, 10.4\right\}$& $\left\{1.1, 1.5\right\}$\\
$(5.0, \left\{0.85, 0.9\right\}, 25)$ & $\left\{4.5, 4.5\right\}$ & $\left\{11.7, 11.9\right\}$& $\left\{1.4, 1.5\right\}$\\
$(5.0, 1.0,  \left\{10, 25, 30\right\})$ & $\left\{5, 4.7, 4.6\right\}$& $\left\{8.5, 12.9, 12.4\right\}$& $\left\{1.1, 1.4, 1.5\right\}$\\
$(5.3, 0.3,  \left\{25, 30, 35\right\})$ & $\left\{5.1, 5.0, 4.8\right\}$& $\left\{14.0, 15.7, 15.7\right\}$& $\left\{0.6, 0.6, 0.7\right\}$\\
$(5.3, 0.6,  \left\{30, 35\right\})$ & $\left\{5.1, 5.0\right\}$& $\left\{15.5, 15.5\right\}$& $\left\{0.6, 0.6\right\}$\\
\hline
\hline
\end{tabular}
\end{table*}

\subsubsection{The clusters survived by sinking}

As shown by Fig.~8 and~9, clusters that manage to reach the bottom of the potential
do not experience significant mass loss thereafter and their structural properties remain
approximately constant. 
A list of the properties of the surviving clusters is provided in Table~2. 
The final mass $M_{c,f}$ and the final projected half light radius $R_{h,c,f}$ are recorded
together with the time it takes for the cluster to sink to the center of the host, $t_{fr}$.
Operationally, this is defined as the time at which the cluster's orbital energy and angular momentum 
imply an instantaneous apocenter $r_{apo}<30$~pc. 
Sinking times span the full available range: from the $t_{fr}\approx15$~Gyr of 
the low-mass cluster with $\log M_{c,i}/M_\odot$ -- which survives because of its
high concentration and perfectly circular orbit, to the much quicker $t_{fr}\approx 0.6$~Gyr
of the models with high initial mass, $\log M_{c,i}/M_\odot> 5$. It is interesting to
notice the good agreement at high masses with the estimates of eqn.~\ref{Chandrasol} and Fig.~2,
when the clusters experience very limited mass mass loss. Sinking times become longer 
than estimated for clusters of initially lower mass, when tidal stripping is more effective.

Figure~10 shows the final masses (left panel) and final projected half light radii (right panel) of all 
surviving clusters, in the three dimensional parameter space. The dependences of the 
final mass are as expected: all the rest being the same, the 
final mass increases with the initial mass, increases with the initial orbital circularity,  
decreases with the initial size. The right panel of the same Figure
shows that large final sizes, $R_{h,c,f}\gtrsim14~$pc, are achieved 
only for high initial masses, $M_{c,i}\gtrsim 10^{5}M_\odot$.
Importantly, I find that no surviving cluster has a final size that is much
larger than its starting one. In all cases $R_{h,c,f}\lesssim r_{c,i}$: if the cluster 
manages to survive, tidal heating has not significantly increased its size.
The final size reflects well the initial size, with clusters that loose high fractions
of their mass and clusters originally more extended experiencing 
stronger reductions in their size. 

In the left panel of Fig.~10, models with a final mass $\log M_{c,f}/M_\odot<3.7$ are highlighted with large red 
circles. These are compatible with the observed mass of the cluster in EriII. Models highlighted with 
blue circles in the same panel have $3.7<\log M_{c,f}/M_\odot<4.4$, compatibly with 
the cluster in AndXXV. Similarly, the right panel highlights models with 
compatible final sizes. No obvious overlap is readily identified.
Two sets of initial conditions are selected in red because of the final mass, these are 
$\left( \log M_{c,i}/M_\odot, j_{c,i}, r_{h,c,i}/{\rm pc}\right) = (3.9, 1.0, 10)$ 
and $\left( \log M_{c,i}/M_\odot, j_{c,i}, r_{h,c,i}/{\rm pc}\right) = (4.7, 0.3, 12.8)$. 
An illustration of the latter is presented in the top panels of Fig.~8. 
Both of these models manage to survive because 
of their small initial sizes and would be disrupted if initially more extended. 
In fact, both have final {\it and} initial sizes that are smaller than observed in 
EriII. Models selected for having compatible final sizes have instead much higher initial mass, $\log M_{c,i}/M_\odot=5$,
and are already extended at formation, $r_{c,i}\geq25$~pc. However, they result in exceedingly high 
final masses. While this does not demonstrate that the cluster in EriII cannot
be a sunk cluster, it does show that clusters with these properties are not a common outcome in 
cuspy haloes of the relevant virial mass. If initial conditions capable of generating a cluster that resembles EriII 
indeed exist, they are very well tuned.

\begin{figure*}
\centering
\includegraphics[width=\textwidth]{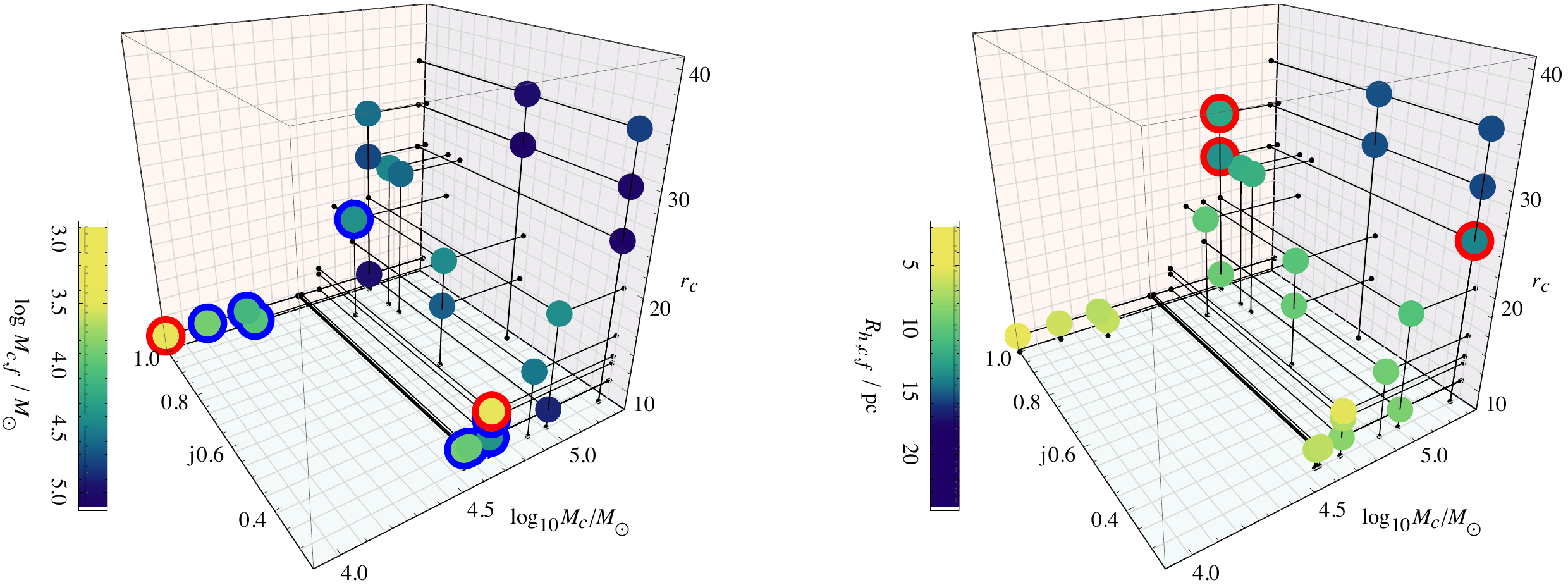}
\caption{The final mass $M_{c,f}$, left panel, and the final projected half-light radius $R_{h,c,f}$ of the sunk clusters.
Models with a final mass compatible with the cluster in EriII, $\log M_{c,f}/M_\odot<3.7$ are highlighted with large red 
circles in the left panel. In the same panel, large blue circles identify models with $3.7<\log M_{c,f}/M_\odot<4.4$,
compatible with the properties of the cluster in AndXXV. In the right panel, similar large circles identify models with 
final projected sizes that are compatible with the cluster in EriII, $12<R_{h,c,f}<14$. No single model with a final 
half-light radius large enough to be compatible with the cluster in AndXXV is found. None of the explored models 
satisfy at the same 
time mass and size requirements to be compatible with observations.}
\end{figure*}

The most promising model for EriII  
$\left( \log M_{c,i}/M_\odot, j_{c,i},\right.$ $\left. r_{h,c,i}/{\rm pc}\right) = (5, 1.0, 30)$, which corresponds to a compatible
final size of $R_{h,c,f}=12.4$~pc, but a high $\log M_{c,f}/M_\odot=4.6$. The evolution of this model is
illustrated in Fig.~9 (top panels), together with the one of the `closeby' model 
$\left( \log M_{c,i}/M_\odot, j_{c,i}, r_{h,c,i}/{\rm pc}\right) = (5, 1.0, 35)$ (bottom panels),
which results in tidal disruption. If the final states of the intervening models 
have similar sizes but gradually decreasing masses, then a cluster similar to the one in EriII could indeed form
from a set of initial conditions within this interval. 
Assuming this is possible, most EriII stars would in fact be stars that formed in the cluster, 
as the initial cluster mass is comparable with the total stellar mass of the galaxy itself. 
Additionally, the tidal friction time scale would be quite short, $t_{fr}\approx1.5~$Gyr.
If indeed existing, this model would belong to a quite small island of the parameter space. 
A very fine tuning is required in order to adjust initial size and initial orbital circularity so for the cluster to 
shed a very high fraction of mass, $\gtrsim95\%$, without however being disrupted. 
Such `viable' island is likely to extend in the direction of higher initial masses, $\log M_{c,i}/M_\odot> 5$, 
but this would not increase its volume significantly. 

The case of AndXXV appears even more desperate: in the entire suite,
no single cluster manages to sink while maintaining a large enough projected half light radius.
All runs have an all too compact $R_{h,c,f}\lesssim16~$pc.
It is much easier to reproduce the final mass of the cluster in AndXXV, but the final
half light radii of these models are approximately factor 2 away from the target size.
This is even more extreme than in the case of EriII. It is possible that 
initially very extended, $r_{c,i}\gtrsim35$~pc, well tuned clusters could produce a 
sunk remnant compatible with AndXXV. These however would likely be initially massive, 
$\log M_{c,i}/M_\odot\gtrsim5.3$, and shed the large majority of their mass. As for EriII, 
this raises a problem of extreme fine tuning.
As explored here, the scenario of sunk clusters in a cuspy halo
does not naturally reproduce the properties of either EriII or AndXXV.

As to the kinematical properties of the sunk clusters, these have low 
line of sight velocities in cuspy haloes, $\delta v_{LOS} \lesssim v_{circ}(r_{stall}\approx5~{\rm pc}) = 1.5~$kms$^{-1}$,
and their internal velocity dispersions are not as inflated as in Section~5.1.1,
as a consequence of their sculpting effect on the very central density cusp,
as described in Sect.~3.3.

\begin{figure*}
\centering
\includegraphics[width=\textwidth]{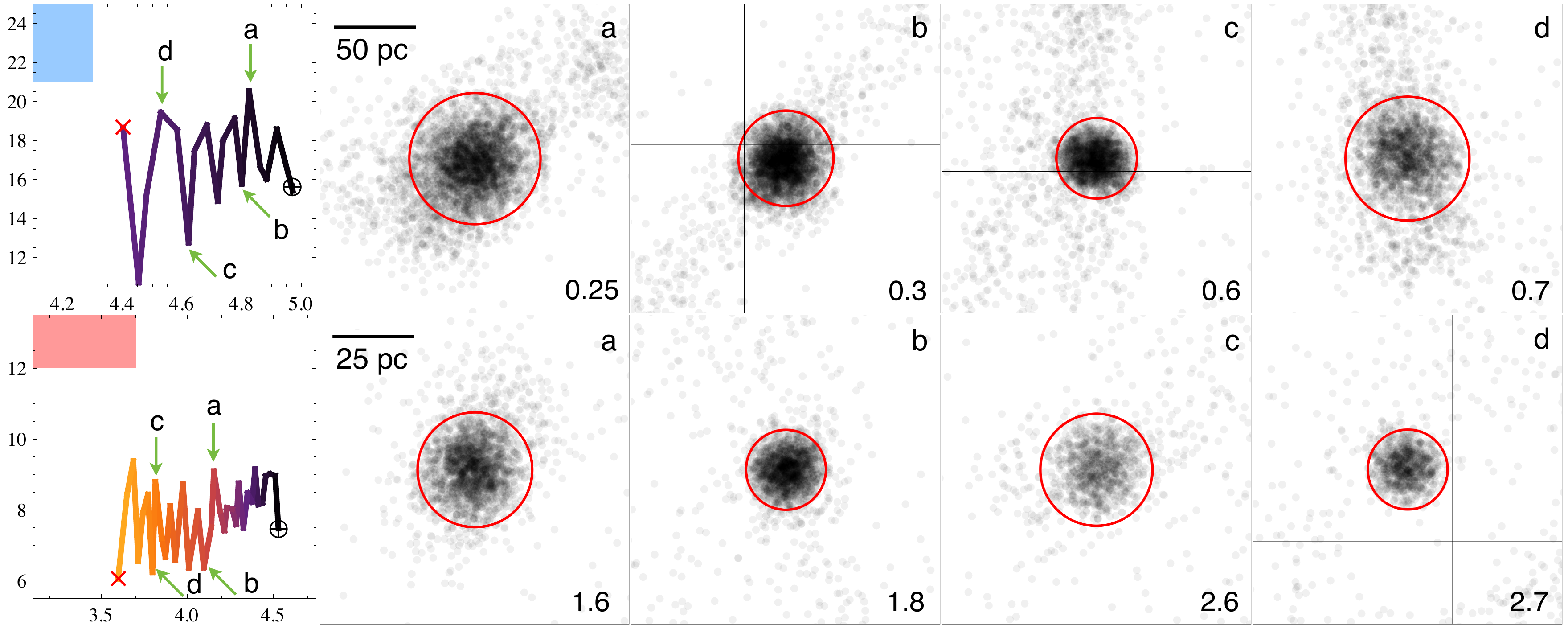}
\caption{An illustration of the oscillation in the size of clusters on eccentric orbits.
Top panels refer to the run $\left( \log M_{c,i}/M_\odot, j_{c,i}, r_{h,c,i}/{\rm pc}\right) = (5, 0.3, 21)$,
bottom panels have $\left( \log M_{c,i}/M_\odot, j_{c,i}, r_{h,c,i}/{\rm pc}\right) = (4.55, 0.3, 10)$. In both rows, the left-most panels
display the evolutionary tracks of the cluster in the plane of cluster
mass (horizontal axis, in units of $\log M_c /M_\odot$) and projected half light radius (vertical axis in units of~pc). 
The regular oscillatory behavior
of the half light radius reflects orbital phase: clusters are contracted at pericenter and expand near apocenter. 
The shaded regions identify the properties of the observed clusters (blue for AndXXV, red for EriII).  
Arrows identify the snapshots illustrated by panels $a - d$. Red circles in panels $a - d$
have a radius of $2R_{h,c}$ at that time; time is displayed in the bottom-right in Gyr. }
\end{figure*}

\subsection{The clusters are in a transitory state}

A last possibility to consider for cuspy haloes is that the clusters are currently in a state that 
departs from dynamical equilibrium, and that the observed structural properties are in fact transitory. 
Clusters on very eccentric orbits are strongly affected by tidal shocking at pericenter: this periodic energy 
injection causes their half light radius to expand and contract on the orbital
timescale. 

Examples of this behavior are shown in Fig.~11, which displays close-ups from the 
evolution of models with $\left( \log M_{c,i}/M_\odot, j_{c,i}, r_{h,c,i}/{\rm pc}\right)=(5,0.3,21)$ in the upper 
panels and $\left( \log M_{c,i}/M_\odot, j_{c,i}, r_{h,c,i}/{\rm pc}\right)=(4.55,0.3,10)$ in the lower panels. 
Both these sets of initial conditions lead to tidal disruption. 
The leftmost panels display the evolutionary tracks of the two clusters in the plane of the instantaneous cluster
mass, on the horizontal axis with units in $\log M_c /M_\odot$, and of the instantaneous projected half light 
radius $R_{h,c}$, on the vertical axis with units in pc. 
Tracks are color-coded by time, and extend between the start of the simulation, marked by a black circle, 
and the last snapshot in which a bound remnant is identified, marked by a red cross. Snapshots 
are available in intervals of $0.05$ and $0.1~$Gyr for the models in the upper and lower panels, respectively.
Both evolutionary tracks show that, while continuously shedding mass, the cluster oscillates between an
expanded and a contracted state, in an a approximately regular fashion. 

Panels $a~-~d$ in the same Figure are centered on the cluster; time is shown in the bottom-right corner
of each panel and red circles have a radius of $2R_{h,c}$ at that time.
The different panels in each row share the same linear scale, illustrated by the black bar in panels $a$,
so that the sizes of the clusters are comparable. In each row, the four panels refer to couples of 
expanded-contracted states, consequent in time, as shown by the green arrows in the left-most panels.
These states correspond to different orbital phases: the clusters are close to pericenter when in their 
contracted state. Then, following the tidal shock, they expand at apocenter. The evolutionary tracks 
of models on circular orbits do not display this behavior and their projected half-light radius decreases in a monotonic 
fashion. This effect is analogous to what recently observed in simulations of dissolving satellite galaxies
on eccentric orbits \citep{AK17}.
It is interesting to note that, when in an expanded state, the internal
velocity dispersion of the cluster remains approximately compatible with the total bound mass, 
i.e. departures from virial equilibrium are not significant.

The shaded regions in the left-most panels identify the properties of the observed clusters, 
in blue for AndXXV in the top row, and in red for EriII in the bottom row. The size oscillations
described above bring models closer to the target properties. I have examined the evolutionary
tracks of all explored models, disrupted and surviving, as in Table~1, but did not identify 
transitory states with the target properties. When in their extended state at apocenter,
initially extended clusters with high initial mass, $\log M_{c,i}/M_\odot\gtrsim5$,
can get `close' to the target region for AndXXV. 
As for the model in the top row of Fig.~11, this happens close in time to full tidal 
disruption. Clusters with this mass sink in $t_{fr}\lesssim1.5~$Gyr, and 
therefore their lifetime is even shorter when they disrupt. 
As a consequence, for this mechanism to be a viable candidate  
for AndXXV, the cluster should either be formed very recently, or have been accreted only very 
recently, both of which appear unlikely.

\section{Cored haloes}

This Section is aimed to provide a numerical confirmation that clusters like 
in EriII or AndXXV would be long lived if hosted by cored haloes, $\alpha \lesssim0.2$. 
As shown in Fig.~1, the tidal radius is a non monotonic function of the orbital radius, making
the inner core regions safe shelters for the clusters. The size of the Roche lobe is
instead minimized at intermediate radii. For haloes compatible with the mass constraint~\ref{massE}, 
the strength of the tidal field is maximized at $r\approx400~$pc.

To explore whether this is threat, I simulate two clusters on circular orbits at $r_{circ}=400~$pc
using a static background potential for the halo, so to ensure the clusters indefinitely remain
where tides are stronger. The initial properties of these clusters 
are as observed, and as in the simulations explored in Sect.~5.1: $\log M_c/M_\odot= 3.7$ and
$R_{h,c}=13$~pc in EriII, $\log M_c/M_\odot= 4.4$ and
$R_{h,c}=25$~pc in AndXXV. Both clusters are populated with $N_c=10^4$ equal mass
particles. Figure~12 shows their mass evolution, in blue and red 
respectively for AndXXV and EriII. Both are extremely long lived:
they experience some mild mass loss within the first Gyr, as they adjust to the surrounding tidal 
field, but their evolution slows down thereafter. The right panels in the same Figure are close ups 
on the clusters themselves at $t=15~$Gyr. As in previous Figures, the red circles have a radius of $2R_{h,c}$.
As suggested by the analytic arguments of Section~3, despite their low densities, 
clusters like those observed in EriII and AndXXV are not threatened by the tidal field
of haloes with $\alpha \lesssim0.2$.

\begin{figure}
\centering
\includegraphics[width=.9\columnwidth]{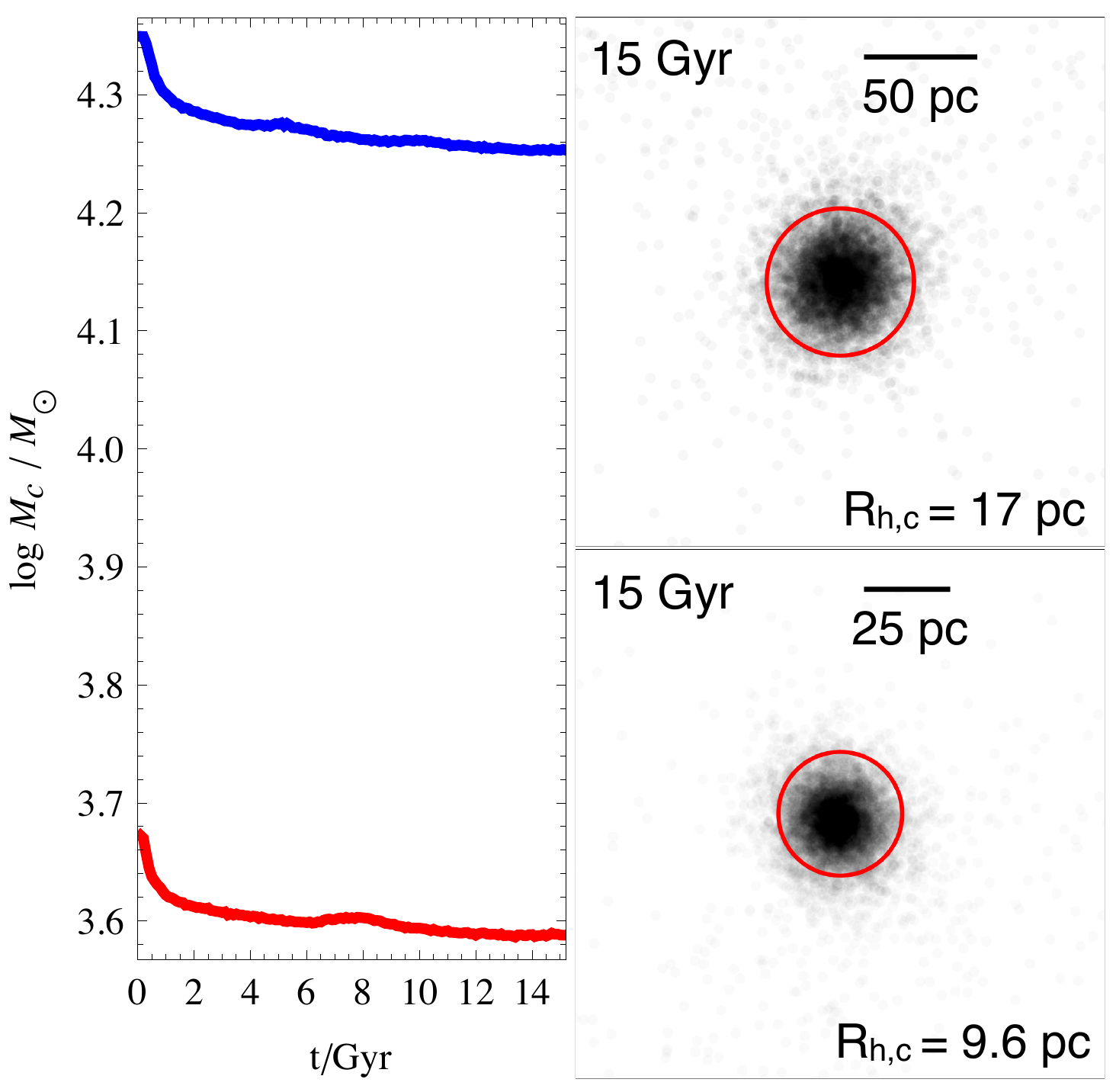}
\caption{Left panel: the mass evolution of clusters as observed in EriII (in red) ad AndXXV (in blue)
when in cored haloes, on circular orbits at the radius where the tidal field is strongest, $r=400~$pc.
Left panels: close ups on the clusters at $t=15~$Gyr. The red circles have a radius of $2R_{h,c}$ at that time.}
\end{figure}

As the cluster in EriII has not shed much mass so far, 
Fig.~2 shows that it is likely to be now close to completing its sinking process.
The cluster in AndXXV should instead already have done so, and reached its stalling radius. 
As shown by panel $c$ in Fig.~2, this makes the observation of small offsets from the dwarf center,
on the scale of 50~pc, quite natural. In both cases, these should be accompanied
by low, though likely non-zero, proper velocities, of the order of the circular velocity
at the stalling radius
\begin{equation}
\delta v_{LOS} \lesssim v_{circ}(r_{stall}\approx50~{\rm pc}) = 1.7~{\rm kms}^{-1} \  .
\label{vstall}
\end{equation}
Finally, as shown by Fig.~6, the DM content of the clusters in this case
should be just a fraction of their stellar content, and their internal velocity dispersions
should therefore be close to what predicted by eqn.~(\ref{sigmac}). 
 
\section{Discussion and Conclusions}

This paper studies the evolution of star clusters in dwarf galaxy haloes, 
concentrating on the regime of clusters with low mass and extended size.
This is aimed to reproduce the evolution and survival to the present day of the two low density 
clusters recently observed in EriII and AndXXV.  A systematic numerical exploration is 
dedicated to the case of cold DM haloes, with a central density cusp with slope $\alpha=1$. 
As suggested by the internal kinematics and by the luminosity of these dwarfs, I concentrate  
on mean cosmological haloes with virial mass of $M_{200}\approx10^9~M_\odot$.
These haloes are found to pose an almost inescapable threat to 
star clusters that are as faint and extended as observed. Haloes 
compatible with the galaxy's internal kinematics but with  
different values for the inner density slope $\alpha$ are found to 
have very similar tidal fields for $\alpha\gtrsim0.2$. 

Clusters as seen in EriII or AndXXV have lifetimes $<1~$Gyr
if orbiting at radii $r\lesssim500$~pc from the center of such haloes:
the probability of observing long lived clusters at small 
galactocentric distances is negligible, at odds with the observations. 
The possibility that the clusters were initially more massive and have sunk
to the center is found to require a very fine tuning. To survive tidal stripping,
a high initial mass is required; however, shedding a high fraction of this mass
without resulting in complete disruption requires well adjusted initial conditions.
Clusters as extended and as faint as observed are not a natural outcome 
in cold DM haloes with the relevant virial mass. 

The most likely scenario offered by cuspy haloes is that the clusters 
formed at rest at the bottom of the potential, where they would not be
subject to the tidal field. {With the present information alone,
this possibility cannot be excluded, but firm predictions can be made. First,} the observed offset from the 
center of the two dwarfs would have to be spurious, caused for 
example by lopsided stellar distributions.  {Second,} this scenario predicts
no offset between the systematic line of sight velocity of the galaxy 
and the one of the cluster. Furthermore, if the process of cluster formation
did not affect the central DM cusp, or preceded the formation of the cusp itself, 
the clusters should be dark matter dominated, resulting in an {observably} inflated internal velocity dispersion.

In turn, in shallow or cored DM haloes, $\alpha \lesssim0.2$, tidal fields do not represent a threat.
In fact, the central regions of the halo are found to be safe shelters: 
the nominal tidal radius has a minimum at a scale of $r/r_s\approx1/4$,
and it increases in the inner core. Clusters like those in EriII and AndXXV
are long lived and free to orbit at any radius in such a tidal field. 
{In this scenario,} dynamical friction timescales are such that the clusters should either have completed their
sinking or be close to reaching their stalling radii: such low orbital energies
make the observed small projected offsets unsurprising. Offsets in the 
line of sight velocity should also be small, {though likely non negligible, with a magnitude
comparable to or smaller than the local rotational velocity}. 
Finally, the internal velocity dispersion of the clusters is compatible with the observed luminosities.

This analysis has of course a series of caveats. Some of these  
are unlikely to seriously affect the conclusions of this work. For example:
\begin{itemize}
\item{the mass range assumed for the hosting halo could be too high, but
it appears unlikely that {\it both}
EriII and AndXXV have virial masses $\log M_{200}/M_\odot\ll 9$.  
Star formation should be strongly hampered in such low mass haloes \citep[see e.g.,][]{Ok08,TS16a}.}
\item{I have ignored the collisional dynamics that takes place inside the clusters, which could have 
caused them to expand after sinking. However, the timescales of collisional processes are exceedingly 
long for such extended and faint clusters \citep[e.g., ][]{MG10,MG11}.}
\item{I have ignored the possible presence of a central massive black hole in the clusters.
This would indeed make them less prone to mass loss, but the black hole mass should be 
higher than the cluster mass itself in order to make a qualitative difference in the picture outlined above.}
\end{itemize}
More important caveats are those connected to the physical conditions of the halo 
at the time of the formation of the clusters. Here, I have assumed the density distribution 
of the host haloes do not evolve in time. This is a good approximation for the halos of dwarf satellite
galaxies, which have early formation times and do not substantially grow at recent times.
Even when isolated, a mean cold DM halo with virial mass of $\log M_{200}/M_\odot= 9$ at $z=0$ already
had a mass of $\log M_{200}/M_\odot\approx8.6$ at $z=3$ \citep[e.g.,][]{Fak10}.
However, if the dynamics that brought to systems like EriII and 
AndXXV took place at much higher redshifts, then this might still be an issue. 
Another relevant caveat is connected to the dynamical friction contributed by the galaxy's gas. 
Note that for this to alter the dynamical friction timescale by factors of order~1
the local gas density should be comparable with the local dark matter density at $z=0$. This
seems high for a dwarf galaxy, but it is difficult to exclude this was possible at high redshift.

Nonetheless, both the possibilities just described are in practice equivalent to the scenario in which the clusters
formed at the center of the cusp. For a cold DM halo that managed to preserve its density cusp 
to the present day other options are strongly disfavored, as the two clusters are exceedingly fragile.
The only realistic scenario is that they formed at the bottom of the potential, and are in fact galactic nuclei. 
This is easily tested by observations of the clusters' kinematics, {which provide a straightforward way 
to distinguish between the remaining viable scenarios.}

{
\begin{itemize}
\item{The easiest case is the one in which either of the clusters is not at rest
in its host, with a proper velocity that is higher than what expected for a sunk cluster in cuspy 
host: $\delta v_{LOS}\gtrsim 1.5~$kms$^{-1}$. This would be proof of a centrally cored halo, $\alpha\lesssim0.2$. }
\item{The case of a negligible proper velocity would be compatible with the scenario of a 
nucleus in a cuspy halo. In fact, a cusp should perhaps be preferred in this case: an entirely negligible velocity
would require better tuning in a cored halo, whose kinematical center is not as well defined. 
The internal velocity dispersion of the cluster would then provide further help,
as the cluster should be dark matter dominated if it is a nucleus in a cuspy host.}
\item{The intermediate case of a non negligible but small proper velocity, $\delta v_{LOS}\lesssim 1.5~$kms$^{-1}$,
would indicate a sunk cluster. Formally, both a cuspy and a cored halo would be possible. In both cases the 
internal velocity dispersion of the cluster would be low, and therefore unhelpful. However, this work shows that an extreme degree of fine 
tuning is required to reproduce the structural properties of the observed clusters in the case of a cuspy halo. A core would be 
a much more natural explanation.}
\end{itemize}}

{This provides a simple and well defined strategy to reliably probe the central density distribution
of two low mass dark matter haloes. Such a measurement would have important implications for our understanding of both
stellar feedback and of the nature of dark matter, independently of the outcome. To start, a clear observational 
confirmation of the long predicted central cusp of cold dark matter haloes is still missing.
On the other hand, the detection of a cored halo would be at least as interesting. 
Stellar feedback is not obviously expected to form large cores in galaxies with such low luminosity.
In fact, that would certainly be surprising for a dSph with $L_V\approx5\times 10^5~L_\odot$ like AndXXV
or a UF with only $L_V\approx6\times 10^4~L_\odot$ like EriII. The presence of a central density core would pose fundamental questions 
on the importance and incidence of stellar feedback in dwarf galaxies at the bottom of the galactic mass scale. Finally, if feedback is indeed
proven to be ineffective at such low luminosities, this measurement would directly point to alternative dark matter candidates.}

\acknowledgments
I would like to thank the referee for a constructive report.

\software{Gadget-2 \citep{VS05}}




\begin{thebibliography}{}


\bibitem[Adams et al.(2014)]{JA14} Adams, J.~J., Simon, J.~D., Fabricius, M.~H., et al.\ 2014, \apj, 789, 63 

\bibitem[Agnello \& Evans(2012)]{AA12} Agnello, A., \& Evans, N.~W.\ 2012, \apjl, 754, L39 

\bibitem[Agnello et al.(2014)]{AA14} Agnello, A., Evans, N.~W., \& Romanowsky, A.~J.\ 2014, \mnras, 442, 3284 

\bibitem[Amorisco \& Evans(2011)]{NA11} Amorisco, N.~C., \& Evans, N.~W.\ 2011, \mnras, 411, 2118 

\bibitem[Amorisco et al.(2014)]{NA14} Amorisco, N.~C., Zavala, J., \& de Boer, T.~J.~L.\ 2014, \apjl, 782, L39 

\bibitem[Amorisco et al.(2014)]{2014Natur.507..335A} Amorisco, N.~C., Evans, N.~W., \& van de Ven, G.\ 2014, \nat, 507, 335 

\bibitem[Amorisco et al.(2013)]{NA13} Amorisco, N.~C., Agnello, A., \& Evans, N.~W.\ 2013, \mnras, 429, L89 

\bibitem[Amorisco \& Evans(2012)]{NA12} Amorisco, N.~C., \& Evans, N.~W.\ 2012, \mnras, 419, 184 

\bibitem[Amorisco(2015)]{NA15} Amorisco, N.~C.\ 2015, \mnras, 450, 575 
\bibitem[Amorisco(2017)]{NA17} Amorisco, N.~C.\ 2017, \mnras, 464, 2882 


\bibitem[Anderson et al.(2014)]{BAO} Anderson, L., Aubourg, {\'E}., Bailey, S., et al.\ 2014, \mnras, 441, 24 

\bibitem[Battaglia et al.(2006)]{GB06} Battaglia, G., Tolstoy, E., Helmi, A., et al.\ 2006, \aap, 459, 423 
\bibitem[Battaglia et al.(2008)]{GB08} Battaglia, G., Helmi, A., Tolstoy, E., et al.\ 2008, \apjl, 681, L13 


\bibitem[Bechtol et al.(2015)]{Be15} Bechtol, K., Drlica-Wagner, A., Balbinot, E., et al.\ 2015, \apj, 807, 50 


\bibitem[Binney \& Tremaine(2008)]{BT08} Binney, J., \& Tremaine, S.\ 2008, Galactic Dynamics: Second Edition, by James Binney and Scott Tremaine.~ISBN 978-0-691-13026-2 (HB).~Published by Princeton University Press, Princeton, NJ USA, 2008.,  

\bibitem[Bode et al.(2001)]{Bo01} Bode, P., Ostriker, J.~P., \& Turok, N.\ 2001, \apj, 556, 93 


\bibitem[Bose et al.(2016)]{SB16} Bose, S., Hellwing, W.~A., Frenk, C.~S., et al.\ 2016, \mnras, 455, 318 

\bibitem[Brandt(2016)]{TB16} Brandt, T.~D.\ 2016, \apjl, 824, L31 


\bibitem[Breddels \& Helmi(2013)]{MB13} Breddels, M.~A., \& Helmi, A.\ 2013, \aap, 558, A35 

{
\bibitem[Brockamp et al.(2014)]{Br14} Brockamp, M., K{\"u}pper, A.~H.~W., Thies, I., Baumgardt, H., \& Kroupa, P.\ 2014, \mnras, 441, 150 
}

\bibitem[Bullock et al.(2001)]{JB01} Bullock, J.~S., Kravtsov, A.~V., \& Weinberg, D.~H.\ 2001, \apj, 548, 33 



\bibitem[Campbell et al.(2016)]{Ca16} Campbell, D.~J.~R., Frenk, C.~S., Jenkins, A., et al.\ 2016, arXiv:1603.04443 

\bibitem[Chandrasekhar(1942)]{Cha42} Chandrasekhar, S.\ 1942, Chicago, Ill., The University of Chicago press [1942],  
\bibitem[Chandrasekhar(1943)]{Cha43} Chandrasekhar, S.\ 1943, \apj, 97, 255 


\bibitem[Cole et al.(2011)]{Co11} Cole, D.~R., Dehnen, W., \& Wilkinson, M.~I.\ 2011, \mnras, 416, 1118 

\bibitem[Cole et al.(2012)]{Co12} Cole, D.~R., Dehnen, W., Read, J.~I., \& Wilkinson, M.~I.\ 2012, \mnras, 426, 601 



\bibitem[Collins et al.(2013)]{MC13} Collins, M.~L.~M., Chapman, S.~C., Rich, R.~M., et al.\ 2013, \apj, 768, 172 
\bibitem[Collins et al.(2014)]{MC14} Collins, M.~L.~M., Chapman, S.~C., Rich, R.~M., et al.\ 2014, \apj, 783, 7 

\bibitem[Cooper et al.(2010)]{AC10} Cooper, A.~P., Cole, S., Frenk, C.~S., et al.\ 2010, \mnras, 406, 744 
\bibitem[Cooper et al.(2013)]{AC13} Cooper, A.~P., D'Souza, R., Kauffmann, G., et al.\ 2013, \mnras, 434, 3348 



\bibitem[Crnojevi{\'c} et al.(2016)]{DC16} Crnojevi{\'c}, D., Sand, D.~J., Zaritsky, D., et al.\ 2016, \apjl, 824, L14 


\bibitem[Cusano et al.(2016)]{FC16} Cusano, F., Garofalo, A., Clementini, G., et al.\ 2016, \apj, 829, 26 



\bibitem[Deason et al.(2014)]{2014ApJ...794..115D} Deason, A., Wetzel, A., \& Garrison-Kimmel, S.\ 2014, \apj, 794, 115 

\bibitem[de Blok \& Bosma(2002)]{dBB02} de Blok, W.~J.~G., \& Bosma, A.\ 2002, \aap, 385, 816 
\bibitem[de Blok et al.(2008)]{dB08} de Blok, W.~J.~G., Walter, F., Brinks, E., et al.\ 2008, \aj, 136, 2648-2719 

\bibitem[Dejonghe(1987)]{HD87} Dejonghe, H.\ 1987, \mnras, 224, 13 


\bibitem[Dekel \& Silk(1986)]{DS86} Dekel, A., \& Silk, J.\ 1986, \apj, 303, 39 

{
\bibitem[Del Popolo \& Pace(2016)]{DP16} Del Popolo, A., \& Pace, F.\ 2016, \apss, 361, 162 
}


\bibitem[den Brok et al.(2014)]{dB14} den Brok, M., Peletier, R.~F., Seth, A., et al.\ 2014, \mnras, 445, 2385 



\bibitem[Di Cintio et al.(2014)]{DC14} Di Cintio, A., Brook, C.~B., Dutton, A.~A., et al.\ 2014, \mnras, 441, 2986 

\bibitem[Diemand et al.(2008)]{JD08} Diemand, J., Kuhlen, M., Madau, P., et al.\ 2008, \nat, 454, 735 

\bibitem[Du et al.(2017)]{Du17} Du, X., Behrens, C., Niemeyer, J.~C., \& Schwabe, B.\ 2017, \prd, 95, 043519 



\bibitem[Dubinski \& Carlberg(1991)]{DC91} Dubinski, J., \& Carlberg, R.~G.\ 1991, \apj, 378, 496 

\bibitem[Eddington(1916)]{Edd16} Eddington, A.~S.\ 1916, \mnras, 76, 572 


\bibitem[Elbert et al.(2015)]{El15} Elbert, O.~D., Bullock, J.~S., Garrison-Kimmel, S., et al.\ 2015, \mnras, 453, 29 

\bibitem[El-Badry et al.(2016)]{KE16} El-Badry, K., Wetzel, A., Geha, M., et al.\ 2016, \apj, 820, 131 

\bibitem[El-Zant et al.(2001)]{EZ01} El-Zant, A., Shlosman, I., \& Hoffman, Y.\ 2001, \apj, 560, 636 


\bibitem[Erkal et al.(2016)]{DE16} Erkal, D., Belokurov, V., Bovy, J., \& Sanders, J.~L.\ 2016, \mnras, 463, 102 


\bibitem[Errani et al.(2015)]{RE15} Errani, R., Pe{\~n}arrubia, J., \& Tormen, G.\ 2015, \mnras, 449, L46 

\bibitem[Fakhouri et al.(2010)]{Fak10} Fakhouri, O., Ma, C.-P., \& Boylan-Kolchin, M.\ 2010, \mnras, 406, 2267 


\bibitem[Fattahi et al.(2016)]{AF16} Fattahi, A., Navarro, J.~F., Sawala, T., et al.\ 2016, arXiv:1607.06479 

\bibitem[Flores \& Primack(1994)]{FP94} Flores, R.~A., \& Primack, J.~R.\ 1994, \apjl, 427, L1 



\bibitem[Frenk \& White(2012)]{FW12} Frenk, C.~S., \& White, S.~D.~M.\ 2012, Annalen der Physik, 524, 507 

{
\bibitem[Galleti et al.(2004)]{Ga04} Galleti, S., Federici, L., Bellazzini, M., Fusi Pecci, F., \& Macrina, S.\ 2004, \aap, 416, 917 
}


\bibitem[Garrison-Kimmel et al.(2017)]{GK17} Garrison-Kimmel, S., Bullock, J.~S., Boylan-Kolchin, M., \& Bardwell, E.\ 2017, \mnras, 464, 3108 



\bibitem[Gibbons et al.(2017)]{2017MNRAS.464..794G} Gibbons, S.~L.~J., Belokurov, V., \& Evans, N.~W.\ 2017, \mnras, 464, 794 

\bibitem[Gieles et al.(2010)]{MG10} Gieles, M., Baumgardt, H., Heggie, D.~C., \& Lamers, H.~J.~G.~L.~M.\ 2010, \mnras, 408, L16 
\bibitem[Gieles et al.(2011)]{MG11} Gieles, M., Heggie, D.~C., \& Zhao, H.\ 2011, \mnras, 413, 2509 


\bibitem[Goerdt et al.(2006)]{Go06} Goerdt, T., Moore, B., Read, J.~I., Stadel, J., \& Zemp, M.\ 2006, \mnras, 368, 1073 
\bibitem[Goerdt et al.(2010)]{Go10} Goerdt, T., Moore, B., Read, J.~I., \& Stadel, J.\ 2010, \apj, 725, 1707 



\bibitem[Governato et al.(2012)]{FG12} Governato, F., Zolotov, A., Pontzen, A., et al.\ 2012, \mnras, 422, 1231 

\bibitem[Hartmann et al.(2011)]{Ha11} Hartmann, M., Debattista, V.~P., Seth, A., Cappellari, M., \& Quinn, T.~R.\ 2011, \mnras, 418, 2697 

{
\bibitem[Hernandez \& Gilmore(1998)]{HG98} Hernandez, X., \& Gilmore, G.\ 1998, \mnras, 297, 517 
}


\bibitem[Hezaveh et al.(2016)]{YH16} Hezaveh, Y.~D., Dalal, N., Marrone, D.~P., et al.\ 2016, \apj, 823, 37 


\bibitem[Hu et al.(2000)]{Hu00} Hu, W., Barkana, R., \& Gruzinov, A.\ 2000, Physical Review Letters, 85, 1158 

\bibitem[Hui et al.(2017)]{Hui17} Hui, L., Ostriker, J.~P., Tremaine, S., \& Witten, E.\ 2017, \prd, 95, 043541 

\bibitem[Hurley \& Mackey(2010)]{HM10} Hurley, J.~R., \& Mackey, A.~D.\ 2010, \mnras, 408, 2353 



\bibitem[Huxor et al.(2011)]{Hux11} Huxor, A.~P., Ferguson, A.~M.~N., Tanvir, N.~R., et al.\ 2011, \mnras, 414, 770 



\bibitem[Ibata et al.(2002)]{RI02} Ibata, R.~A., Lewis, G.~F., Irwin, M.~J., \& Quinn, T.\ 2002, \mnras, 332, 915 

\bibitem[Inoue(2009)]{In09} Inoue, S.\ 2009, \mnras, 397, 709 



\bibitem[Jethwa et al.(2016)]{PJ16} Jethwa, P., Belokurov, V., \& Erkal, D.\ 2016, arXiv:1612.07834 



\bibitem[Johnston et al.(2002)]{Yo11} Johnston, K.~V., Spergel, D.~N., \& Haydn, C.\ 2002, \apj, 570, 656 




\bibitem[Kleyna et al.(2003)]{Kl03} Kleyna, J.~T., Wilkinson, M.~I., Gilmore, G., \& Evans, N.~W.\ 2003, \apjl, 588, L21 

\bibitem[Kochanek \& Dalal(2004)]{KD04} Kochanek, C.~S., \& Dalal, N.\ 2004, \apj, 610, 69 

\bibitem[Koposov et al.(2015)]{SK15} Koposov, S.~E., Belokurov, V., Torrealba, G., \& Evans, N.~W.\ 2015, \apj, 805, 130 


\bibitem[Kordopatis et al.(2016)]{GK16} Kordopatis, G., Amorisco, N.~C., Evans, N.~W., Gilmore, G., \& Koposov, S.~E.\ 2016, \mnras, 457, 1299 

\bibitem[Koushiappas \& Loeb(2017)]{KL17} Koushiappas, S.~M., \& Loeb, A.\ 2017, arXiv:1704.01668 


\bibitem[K{\"u}pper et al.(2017)]{AK17} K{\"u}pper, A.~H.~W., Johnston, K.~V., Mieske, S., Collins, M.~L.~M., \& Tollerud, E.~J.\ 2017, \apj, 834, 112

\bibitem[Li et al.(2017)]{Li17} Li, T.~S., Simon, J.~D., Drlica-Wagner, A., et al.\ 2017, \apj, 838, 8 


\bibitem[Lin \& Loeb(2016)]{LA16} Lin, H.~W., \& Loeb, A.\ 2016, \jcap, 3, 009 

\bibitem[Lotz et al.(2001)]{JL01} Lotz, J.~M., Telford, R., Ferguson, H.~C., et al.\ 2001, \apj, 552, 572 



\bibitem[Lovell et al.(2014)]{ML14} Lovell, M.~R., Frenk, C.~S., Eke, V.~R., et al.\ 2014, \mnras, 439, 300 

\bibitem[Ludlow et al.(2016)]{AL16} Ludlow, A.~D., Bose, S., Angulo, R.~E., et al.\ 2016, \mnras, 460, 1214 

\bibitem[Huxor et al.(2013)]{Hux13} Huxor, A.~P., Ferguson, A.~M.~N., Veljanoski, J., Mackey, A.~D., \& Tanvir, N.~R.\ 2013, \mnras, 429, 1039 

{
\bibitem[Ma \& Boylan-Kolchin(2004)]{MB04} Ma, C.-P., \& Boylan-Kolchin, M.\ 2004, Physical Review Letters, 93, 021301 
}


\bibitem[Mackey et al.(2010)]{AM10} Mackey, A.~D., Huxor, A.~P., Ferguson, A.~M.~N., et al.\ 2010, \apjl, 717, L11 


\bibitem[Martin et al.(2016)]{NM16} Martin, N.~F., Ibata, R.~A., Lewis, G.~F., et al.\ 2016, \apj, 833, 167 


\bibitem[Mashchenko et al.(2006)]{Ma06} Mashchenko, S., Couchman, H.~M.~P., \& Wadsley, J.\ 2006, \nat, 442, 539 

\bibitem[McConnachie(2012)]{MC12} McConnachie, A.~W.\ 2012, \aj, 144, 4 


\bibitem[Menci et al.(2012)]{Men12} Menci, N., Fiore, F., \& Lamastra, A.\ 2012, \mnras, 421, 2384 

\bibitem[Mocz \& Succi(2015)]{PM15} Mocz, P., \& Succi, S.\ 2015, \pre, 91, 053304 



\bibitem[Navarro et al.(1996)]{NFW96} Navarro, J.~F., Frenk, C.~S., \& White, S.~D.~M.\ 1996, \apj, 462, 563 

\bibitem[Navarro et al.(1996)]{JN96} Navarro, J.~F., Eke, V.~R., \& Frenk, C.~S.\ 1996, \mnras, 283, L72 

\bibitem[Nipoti \& Binney(2015)]{NB15} Nipoti, C., \& Binney, J.\ 2015, \mnras, 446, 1820 



\bibitem[Oh et al.(2011)]{Oh11} Oh, S.-H., de Blok, W.~J.~G., Brinks, E., Walter, F., \& Kennicutt, R.~C., Jr.\ 2011, \aj, 141, 193 

\bibitem[Oh \& Lin(2000)]{Oh00} Oh, K.~S., \& Lin, D.~N.~C.\ 2000, \apj, 543, 620 

\bibitem[Okamoto et al.(2008)]{Ok08} Okamoto, T., Gao, L., \& Theuns, T.\ 2008, \mnras, 390, 920 


\bibitem[Oman et al.(2015)]{KO15} Oman, K.~A., Navarro, J.~F., Fattahi, A., et al.\ 2015, \mnras, 452, 3650 



\bibitem[O{\~n}orbe et al.(2015)]{JO15} O{\~n}orbe, J., Boylan-Kolchin, M., Bullock, J.~S., et al.\ 2015, \mnras, 454, 2092 

{
\bibitem[Pe{\~n}arrubia et al.(2009)]{JP09} Pe{\~n}arrubia, J., Walker, M.~G., \& Gilmore, G.\ 2009, \mnras, 399, 1275 
}

\bibitem[Pe{\~n}arrubia et al.(2012)]{JP12} Pe{\~n}arrubia, J., Pontzen, A., Walker, M.~G., \& Koposov, S.~E.\ 2012, \apjl, 759, L42 
\bibitem[Pe{\~n}arrubia et al.(2016)]{JP16} Pe{\~n}arrubia, J., Ludlow, A.~D., Chanam{\'e}, J., \& Walker, M.~G.\ 2016, \mnras, 461, L72 

\bibitem[Persic \& Salucci(1991)]{PS91} Persic, M., \& Salucci, P.\ 1991, \apj, 368, 60 

\bibitem[Petts et al.(2016)]{Pe16} Petts, J.~A., Read, J.~I., \& Gualandris, A.\ 2016, \mnras, 463, 858 


\bibitem[Planck Collaboration et al.(2014)]{Planck} Planck Collaboration, Ade, P.~A.~R., Aghanim, N., et al.\ 2014, \aap, 571, A16 

\bibitem[Plummer(1911)]{Pl11} Plummer, H.~C.\ 1911, \mnras, 71, 460 



\bibitem[Pontzen \& Governato(2014)]{PG14} Pontzen, A., \& Governato, F.\ 2014, \nat, 506, 171 

\bibitem[Press et al.(1990)]{Pr90} Press, W.~H., Ryden, B.~S., \& Spergel, D.~N.\ 1990, Physical Review Letters, 64, 1084 


\bibitem[Read et al.(2006)]{JR06} Read, J.~I., Goerdt, T., Moore, B., et al.\ 2006, \mnras, 373, 1451 
\bibitem[Read et al.(2016)]{JR16} Read, J.~I., Agertz, O., \& Collins, M.~L.~M.\ 2016, \mnras, 459, 2573 

\bibitem[Renaud et al.(2011)]{FR11} Renaud, F., Gieles, M., \& Boily, C.~M.\ 2011, \mnras, 418, 759 

\bibitem[Richardson et al.(2011)]{Ri11} Richardson, J.~C., Irwin, M.~J., McConnachie, A.~W., et al.\ 2011, \apj, 732, 76 


\bibitem[Richardson \& Fairbairn(2014)]{RF14} Richardson, T., \& Fairbairn, M.\ 2014, \mnras, 441, 1584 

\bibitem[S{\'a}nchez-Salcedo et al.(2006)]{SS06} S{\'a}nchez-Salcedo, F.~J., Reyes-Iturbide, J., \& Hernandez, X.\ 2006, \mnras, 370, 1829 
\bibitem[S{\'a}nchez-Salcedo \& Lora(2010)]{SS10} S{\'a}nchez-Salcedo, F.~J., \& Lora, V.\ 2010, \mnras, 407, 1135 

\bibitem[Sawala et al.(2016)]{TS16a} Sawala, T., Frenk, C.~S., Fattahi, A., et al.\ 2016, \mnras, 456, 85 
\bibitem[Sawala et al.(2016)]{TS16} Sawala, T., Frenk, C.~S., Fattahi, A., et al.\ 2016, \mnras, 457, 1931 



\bibitem[Schive et al.(2014)]{HS14} Schive, H.-Y., Chiueh, T., \& Broadhurst, T.\ 2014, Nature Physics, 10, 496 
\bibitem[Schive et al.(2016)]{HS16} Schive, H.-Y., Chiueh, T., Broadhurst, T., \& Huang, K.-W.\ 2016, \apj, 818, 89 


\bibitem[Spergel \& Steinhardt(2000)]{SS00} Spergel, D.~N., \& Steinhardt, P.~J.\ 2000, Physical Review Letters, 84, 3760 

\bibitem[Springel(2005)]{VS05} Springel, V.\ 2005, \mnras, 364, 1105 

\bibitem[Springel et al.(2008)]{VS08} Springel, V., Wang, J., Vogelsberger, M., et al.\ 2008, \mnras, 391, 1685 

\bibitem[Strigari et al.(2017)]{St17} Strigari, L.~E., Frenk, C.~S., \& White, S.~D.~M.\ 2017, \apj, 838, 123 


\bibitem[Tolstoy et al.(2004)]{ET04} Tolstoy, E., Irwin, M.~J., Helmi, A., et al.\ 2004, \apjl, 617, L119 

\bibitem[Tremaine et al.(1975)]{Tre75} Tremaine, S.~D., Ostriker, J.~P., \& Spitzer, L., Jr.\ 1975, \apj, 196, 407 



\bibitem[Vegetti et al.(2014)]{SV14} Vegetti, S., Koopmans, L.~V.~E., Auger, M.~W., Treu, T., \& Bolton, A.~S.\ 2014, \mnras, 442, 2017 

{
\bibitem[Veljanoski et al.(2013)]{Ve13} Veljanoski, J., Ferguson, A.~M.~N., Mackey, A.~D., et al.\ 2013, \apjl, 768, L33 
}


\bibitem[Vogelsberger et al.(2012)]{VZ12} Vogelsberger, M., Zavala, J., \& Loeb, A.\ 2012, \mnras, 423, 3740 


\bibitem[Walker et al.(2009)]{MW09a} Walker, M.~G., Mateo, M., \& Olszewski, E.~W.\ 2009, \aj, 137, 3100 

\bibitem[Walker et al.(2009)]{MW09} Walker, M.~G., Mateo, M., Olszewski, E.~W., et al.\ 2009, \apj, 704, 1274 

\bibitem[Walker \& Pe{\~n}arrubia(2011)]{WP11} Walker, M.~G., \& Pe{\~n}arrubia, J.\ 2011, \apj, 742, 20 

\bibitem[Widrow(2000)]{LW00} Widrow, L.~M.\ 2000, \apjs, 131, 39 


\bibitem[Wolf et al.(2010)]{JW10} Wolf, J., Martinez, G.~D., Bullock, J.~S., et al.\ 2010, \mnras, 406, 1220 

\bibitem[Zaritsky et al.(2016)]{DZ16} Zaritsky, D., Crnojevi{\'c}, D., \& Sand, D.~J.\ 2016, \apjl, 826, L9 


\bibitem[Zavala et al.(2013)]{JZ13} Zavala, J., Vogelsberger, M., \& Walker, M.~G.\ 2013, \mnras, 431, L20 


\bibitem[Zhu et al.(2016)]{Zhu16} Zhu, L., van de Ven, G., Watkins, L.~L., \& Posti, L.\ 2016, \mnras, 463, 1117 

\bibitem[Zolotov et al.(2012)]{AZ12} Zolotov, A., Brooks, A.~M., Willman, B., et al.\ 2012, \apj, 761, 71 



\end{thebibliography}
\end{document}